\documentclass[manuscript]{acmart}

\AtBeginDocument{%
  \providecommand\BibTeX{{%
    \normalfont B\kern-0.5em{\scshape i\kern-0.25em b}\kern-0.8em\TeX}}}

\setcopyright{acmcopyright}
\acmJournal{CSUR}
\acmYear{2025}

\usepackage{lineno}
\usepackage{svg}
\usepackage{amsmath}  
\usepackage{tabularx}
\usepackage{array,makecell}

\usepackage[flushleft]{threeparttable} %
\usepackage{booktabs,caption}
\usepackage{booktabs}
\usepackage{graphicx}
\usepackage{rotating} 
\usepackage{tabularray} 
\usepackage{soul}
\usepackage{subcaption}

\usepackage[font=small,skip=0pt]{caption}
\usepackage{amsmath} 
\usepackage{fixltx2e}

\usepackage{mathtools}
\usepackage{blkarray, bigstrut}
\usepackage{gauss}
\usepackage{multirow}
\usepackage{multicol}
\usepackage{longtable} 
\usepackage{enumitem} 
\usepackage{setspace}
\usepackage{microtype}

\author{Zahra Mousavi}
\email{seyedehzahra.mosavi@adelaide.edu.au}
\affiliation{
  \institution{Centre for Research on Engineering Software Technologies (CREST) \& Adelaide University, Cyber Security CRC, CSIRO/Data61}
  \country{Australia}
}

\author{Chadni Islam}
\email{c.islam@ecu.edu.au}
\affiliation{%
  \institution{Edith Cowan University}
  \country{Australia}
}

\author{Muhammad Ali Babar}
\affiliation{%
  \institution{Centre for Research on Engineering Software Technologies (CREST) \& Adelaide University}
  \country{Australia}
}

\author{Alsharif Abuadbba}
\author{Kristen Moore}
\affiliation{%
  \institution{CSIRO/Data61}
    \country{Australia}
}

\newcommand*\subtxt[1]{_{\textnormal{#1}}}
\DeclareRobustCommand\_{\ifmmode\expandafter\subtxt\else\textunderscore\fi}

\usepackage{multicol}
\usepackage{listings}
\usepackage{array}
\usepackage{tikz}

\newcolumntype{C}[1]{>{\centering\arraybackslash}p{#1}}

\definecolor{codegreen}{rgb}{0.0, 0.5, 0.0}        
\definecolor{codegray}{rgb}{0.4, 0.4, 0.4}          
\definecolor{codepurple}{rgb}{0.58, 0.0, 0.82}      
\definecolor{codeblue}{rgb}{0.0, 0.0, 0.8}          
\definecolor{codeorange}{rgb}{0.8, 0.3, 0.0}        
\definecolor{backcolour}{rgb}{0.97, 0.97, 0.97}     
\definecolor{lightcoral}{rgb}{0.97, 0.7, 0.7}  
\definecolor{lightcyan}{rgb}{0.60, 1.00, 1.00} 

\lstdefinestyle{mystyle}{
  backgroundcolor=\color{backcolour},
  commentstyle=\color{codegreen}\itshape,
  keywordstyle=\color{codeblue}\bfseries,
  numberstyle=\tiny\color{codegray},
  stringstyle=\color{codepurple},
  basicstyle=\ttfamily\tiny\linespread{1.1}, 
  breakatwhitespace=false,        
  breaklines=true,                
  captionpos=b,                    
  keepspaces=true,                
  numbers=left,                    
  numbersep=4pt,                  
  showspaces=false,                
  showstringspaces=false,
  showtabs=false,                  
  tabsize=2,
  frame=lines
}
\lstdefinelanguage{JavaScript}{
  keywords=[1]{break, case, catch, continue, debugger, default, delete, do, else, finally, for, function, if, in, instanceof, new, return, switch, this, throw, try, typeof, var, void, while, with},
  keywordstyle=[1]\color{codeblue}\bfseries,
  keywords=[2]{class, const, enum, export, extends, import, super, implements, interface, let, package, private, protected, public, static, yield},
  keywordstyle=[2]\color{codeorange}\bfseries,
  keywords=[3]{false, true, null, undefined, NaN, Infinity},
  keywordstyle=[3]\color{codepurple}\bfseries,
  sensitive=false,
  comment=[l]{//},
  morecomment=[s]{/*}{*/},
  commentstyle=\color{codegreen}\ttfamily\itshape,
  stringstyle=\color{codepurple}\ttfamily,
  morestring=[b]',
  morestring=[b]",
  literate=
    *{0}{{{\color{codepurple}0}}}{1}
    {1}{{{\color{codepurple}1}}}{1}
    {2}{{{\color{codepurple}2}}}{1}
    {3}{{{\color{codepurple}3}}}{1}
    {4}{{{\color{codepurple}4}}}{1}
    {5}{{{\color{codepurple}5}}}{1}
    {6}{{{\color{codepurple}6}}}{1}
    {7}{{{\color{codepurple}7}}}{1}
    {8}{{{\color{codepurple}8}}}{1}
    {9}{{{\color{codepurple}9}}}{1},
  basicstyle=\ttfamily\tiny\linespread{1.1},
  breakatwhitespace=false,
  breaklines=true,
  captionpos=b,
  keepspaces=true,
  numbers=left,
  numbersep=4pt,
  showspaces=false,
  showstringspaces=false,
  showtabs=false,
  tabsize=2,
  frame=lines
}

\begin{document}

\title{Detecting Misuse of Security APIs: A Systematic Review}

\begin{abstract}

Security Application Programming Interfaces (APIs) are crucial for ensuring software security. However, their misuse introduces vulnerabilities, potentially leading to severe data breaches and substantial financial loss. Complex API design, inadequate documentation, and insufficient security training often lead to unintentional misuse by developers. The software security community has devised and evaluated several approaches to detecting security API misuse to help developers and organizations. This study rigorously reviews the literature on detecting misuse of security APIs to gain a comprehensive understanding of this critical domain. Our goal is to identify and analyze security API misuses, the detection approaches developed, and the evaluation methodologies employed along with the open research avenues to advance the state-of-the-art in this area. Employing the systematic literature review (SLR) methodology, we analyzed 69 research papers. Our review has yielded (a) identification of 6 security API types; (b) classification of 30 distinct misuses; (c) categorization of detection techniques into heuristic-based and ML-based approaches; and (d) identification of 10 performance measures and 9 evaluation benchmarks. The review reveals a lack of coverage of detection approaches in several areas. We recommend that future efforts focus on aligning security API development with developers' needs and advancing standardized evaluation methods for detection technologies.

\end{abstract}



\maketitle


\raggedbottom 

\section{Introduction}
Security Application Programming Interfaces (APIs) are integral to modern software development that serves billions of users through web or mobile apps. They offer developers specific functionalities, such as data encryption or access control, to address security concerns. Cryptography and SSL/TLS APIs, for example, are widely used to ensure data confidentiality and secure communications ~\cite{georgiev2012most}.
Although security APIs are highly beneficial for ensuring software security, their incorrect use, known as misuse, inadvertently leads to software vulnerabilities, posing a significant risk to overall system security and potentially exposing millions of users to sensitive data breaches and financial losses~\cite{georgiev2012most, egele2013empirical, rahaman2019cryptoguard,bianchi2018broken, al2019oauthlint, kruger2019crysl}. 

\begin{figure}[t!]
  \centering
  \includegraphics[width=.9\columnwidth]{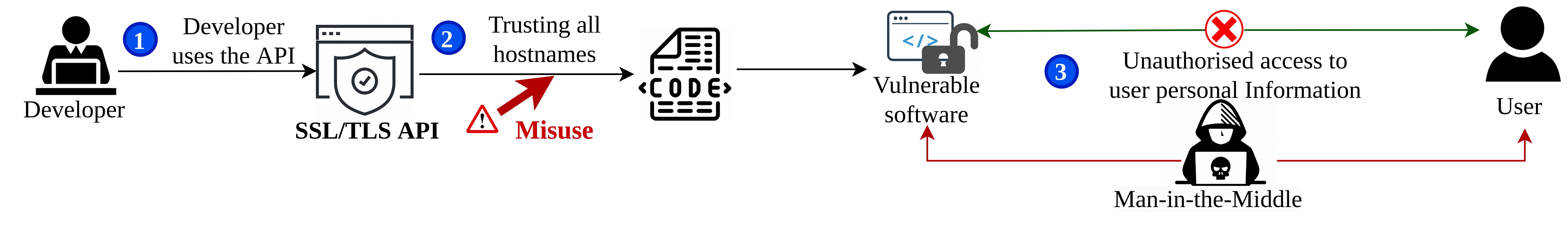}
  \caption{A misuse of SSL/TLS API leading to the leakage of user personal information}
  \label{fig:misuse}  
\end{figure}

Figure~\ref{fig:misuse} illustrates an SSL/TLS API misuse, where a developer uses the API to establish a secure connection with a server (Step \textcircled{\scriptsize 1})  but chooses to \textit{trust all server hostnames} (Step \textcircled{\scriptsize 2}). By exploiting the misuse, a malicious actor can impersonate a valid server, intercept the communication between a user and the application, and obtain unauthorized access to the user's personal information (Step \textcircled{\scriptsize 3}). Many developers do not fully understand the implications of \textit{trusting all hostnames} while using an SSL/TLS API~\cite{georgiev2012most}. The complex design of security APIs, poor documentation, and inadequate training significantly hamper developers' understanding of security APIs and contribute to the widespread misuse of these APIs~\cite{green2016developers}. A comprehensive study of ten thousand Android applications by Krüger et al.~\cite{kruger2019crysl} revealed that nearly 95\% of the applications contain at least one cryptography API misuse. Similarly, an analysis of over two thousand open-source Java projects on GitHub found that 72\% of them suffer from at least one cryptography API misuse~\cite{hazhirpasand2019impact}.


Given the increasing realization of the potentially devastating consequences of misuse of security APIs, there has been significant interest in devising and evaluating effective and efficient approaches to detecting security API misuses. 
Nevertheless, the relevant literature is dispersed and lacks coherent analysis and synthesis. This gap underscores the need for a systematic survey that could assist both researchers and practitioners in gaining a comprehensive understanding of state-of-the-art approaches in this field. Therefore, we aim to conduct a thorough and holistic analysis of the literature specifically concerning the detection of security API misuse.
We leverage the Systematic Literature Review (SLR) methodology and focus on four key dimensions: \textit{the security APIs studied, the types of misuses reported, the detection techniques proposed}, and \textit{the evaluation methods employed} that require attention. This leads to the formation of four primary research questions:
\textbf{(RQ1)} \textit{What \textbf{security APIs} are commonly analyzed in the context of misuse detection across software applications?} \textbf{(RQ2)} \textit{What security API \textbf{misuses} are commonly investigated in the literature for detection across software applications?} \textbf{(RQ3)} \textit{What \textbf{techniques} are commonly employed to detect security API misuses?} \textbf{(RQ4)} \textit{What \textbf{evaluation strategies} are commonly utilized to assess the performance of misuse detection techniques?} Our SLR, covering~69 peer-reviewed studies, yields the following main \textbf{contributions}:

{
\begin{itemize}
    \item A large-scale survey of the literature on misuse detection for security APIs using a systematic review method. 
    \item An in-depth discussion on the security APIs studied from misuse perspective and their misuses.
    \item A taxonomic analysis of the existing approaches to detecting misuse of security APIs.
    \item A critical rundown of the strategies, benchmarks, and metrics used for evaluating the proposed approaches.
    \item A set of open issues that can form the future research agenda for enhancing and evaluating security API use. 
\end{itemize}}

The remainder of this paper is structured as follows. 
Section~\ref{sec:preli} provides an overview of security APIs and their misuses. Section~\ref{sec:methodology} outlines the SLR methodology. Findings on security APIs and misuses are presented in Sections~\ref{sec:APIs} and~\ref{sec:Misuses}, respectively. Section~\ref{sec:tech} analyzes misuse detection techniques, and Section~\ref{sec:eval} covers evaluation strategies. Section~\ref{subsec:open_issues} discusses open issues and future directions. Section~\ref{sec:threats} addresses threats to validity. The paper concludes in Section~\ref{sec: concl}.


\section{Preliminaries}\label{sec:preli}
This section presents an overview of security APIs and potential misuses that developers may make while using them.

\subsection{Security API}\label{subsec:overview_API}
An API is essentially a set of programming instructions that allows software components to interact with each other, making it easier for developers to build complex software systems~\cite{lamothe2021systematic}. 
Security APIs are a subset of APIs that provide developers with security functionalities, such as confidentiality, data integrity, authentication, and authorization~\cite{gorski2016towards}. Confidentiality is the process of protecting sensitive information from unauthorized disclosure~\cite{whitman2013management}. Data encryption, provided by cryptography APIs, is a primary means to ensure the confidentiality of sensitive data.  
Data integrity mechanisms ensure the accuracy, trustworthiness, and validity of data by protecting it from unauthorized changes throughout its life cycle. Cryptography and SSL/TLS APIs are among the popular security APIs that provide security functionalities to ensure data integrity~\cite{kruger2019crysl}. 
Authentication is the process of verifying the identity of legitimate users or systems before granting data access~\cite{whitman2013management}, while authorization specifies access rights and privileges to resources~\cite{whitman2013management}. OAuth APIs~\cite{hardt2012rfc} are widely used for both authentication and authorization purposes, enabling users to grant access to their resources and data to third-party applications without exposing their credentials.

\subsection{Misuse of Security APIs}
APIs function correctly only when specific constraints on their inputs, outputs, and invocation context are met, as outlined in the API specification. Any deviation from these specifications, known as misuse, can lead to various problems such as performance degradation, compatibility issues, and unexpected behavior. However, misusing security APIs can have far more severe consequences, including exposing confidential data and putting entire systems at risk.
Moreover, misusing security APIs can lead to non-compliance with data protection regulations such as GDPR~\cite{krzysztofek2018gdpr} and HIPAA~\cite{hipaa}, potentially resulting in hefty fines and severe damage to a company’s reputation. To mitigate these risks, developers must strictly adhere to the latest specifications when using security APIs. For instance, secure communication via an SSL/TLS API requires a client to verify the server's hostname. Figure~\ref{fig:misuse} illustrates a violation of this specification, demonstrating how neglecting hostname verification in the SSL/TLS API can lead to a Man-in-the-Middle attack. In this scenario, attackers can impersonate a legitimate server, intercepting communication between the user and the application, thereby gaining unauthorized access to the user's personal information.

As digital transformation accelerates, the use of APIs, including security APIs, is growing. The increasing interconnectivity of systems and the proliferation of IoT devices expand the attack surface, making the correct use of security APIs more critical than ever. However, the research reveals a concerning trend of widespread misuse of these APIs among developers.~\cite{kruger2019crysl, hazhirpasand2019impact}, which can be attributed to several factors. Firstly, the complex and nuanced operation of security APIs coupled with the inherent complexity of API design can hinder developers' understanding and correct implementation~\cite{green2016developers}. Security API use is further complicated by poorly written documentation, often lacking clear usage examples, and sometimes including insecure code examples. For instance, misuses were identified in code samples within the documentation of an OAuth API, which can mislead developers who copy them without understanding security implications~\cite {shernan2015more}. Additionally, developers often use unreliable forum posts for guidance, which can propagate instances of misuse~\cite {fischer2017stack}. Lastly, the lack of cybersecurity training, coupled with the evolving threat landscape surrounding security APIs, makes it challenging for developers to stay current on the latest best practices for security APIs~\cite{assal2018security}.

\section{Research methodology}\label{sec:methodology}

We conducted a Systematic Literature Review (SLR) to gain insight into misuse detection approaches for security APIs. SLR is broadly adopted as a research methodology in Evidence-Based Software Engineering~\cite{kitchenham2004evidence} as it provides a reliable, rigorous, and auditable technique for assessing and interpreting a research topic~\cite{slr_guidelines}. We followed the SLR guideline provided by Kitchenham et al.~\cite{slr_guidelines}.
The steps of our review protocol are elaborated in Subsections~\ref{subsec:search_strategy}-~\ref{subsec:data_extract_synthesis}. 

\subsection{Search Strategy}\label{subsec:search_strategy}

We followed the guidelines provided by Kitchenham et al.~\cite{slr_guidelines} to develop our search strategy, ensuring that we obtain the highest number of relevant studies. The search strategy is comprised of the following steps.

\subsubsection{Search Method}
We applied an automated database search method~\cite{slr_guidelines} to digital search engines and databases to obtain relevant studies. We used Scopus Digital Library (DL) as the primary source, which is the largest academic literature database, indexing over 5,000 publishers worldwide, including relevant sources like Elsevier and Springer~\cite{slr_triet_vul, roland}. To complement Scopus results, we also used the two prominent academic DLs -- IEEE Xplore and ACM DL ~\cite{digital_lib}.

\subsubsection{Search String}
We crafted a comprehensive search string following the guidelines presented by Kitchenham et al.~\cite{slr_guidelines}. We initiated our search using four main keywords: ``security'', ``API'', ``misuse'', and ``detection''. To broaden our search, we also considered synonyms for these terms. We reviewed titles, abstracts, and keywords from some relevant papers to ensure we captured associated  synonyms. Synonyms that returned an excess of irrelevant results, such as ``flaw'', were excluded from our search string. We conducted a series of pilot searches to ensure the inclusion of relevant papers that we were already aware of. Ultimately, we organized the keywords and their pertinent synonyms into four categories, which are shown in Table~\ref{tab:str}.  We used the union (AND) of the categories to conduct searches in titles, abstracts, and keywords of papers on Scopus, IEEE Xplorer, and ACM DL.

\begin{table}[h]
\centering
\caption{Categories of key terms used for defining search string}

\label{tab:str}
\small
\renewcommand{\arraystretch}{.9}
\begin{tabularx}{0.82\textwidth}{ll}

\hline
\textbf{Category} & \textbf{Synonyms and Relevant Terms}\\
\hline

\textbf{Security} &  \textit{secur*  OR  crypto*}   \\  
\textbf{API} & \textit{api  OR  librar*  OR  interface}   \\   
\textbf{Misuse} & \textit{misuse  OR  incorrect  OR  insecure  OR  vulnerabilit* }  \\  
\textbf{Detection} & \textit{detect*  OR  ``static analysis''  OR  ``dynamic analysis''  OR  ``program analysis''  OR  ``code analysis''}  \\
\hline 
\end{tabularx}
\end{table}

\subsection{Inclusion-Exclusion Criteria} \label{subsec:quality_assess}
We applied the inclusion and exclusion criteria outlined in Table~\ref{tab:IE} to filter the retrieved studies. We included studies on detecting misuse of security APIs in line with our research objectives and RQs (I1). The criteria were iteratively refined during the selection process to ensure an accurate selection of relevant papers. For instance, we excluded studies that focused on vulnerabilities in the internal design or implementation of security APIs (E1) or targeted at detecting misuses of generic APIs (E2). We also introduced E3 and E4 to exclude publications, such as short papers, that lack sufficient information to fully address the RQs. Moreover, we adopted a venue assessment criterion (I2), similar to methods used in previous studies~\cite{roland, sabir2021machine, sarker2024multi}, to include only high-quality papers. This criterion involved including only papers published in ranked conferences according to the CORE ranking\footnote{\url{http://portal.core.edu.au/conf-ranks}}, and ranked journals according to the Scimago database\footnote{\url{https://www.scimagojr.com/journalrank.php}}. Both databases employ meticulous and comprehensive evaluation methodologies, considering various factors to assess the quality and impact of venues. By utilizing these databases, we were able to identify high-quality papers effectively.

\begin{table}[h]
\centering
\caption{Inclusion and Exclusion Criteria}

\label{tab:IE}
\small
\begin{tabularx}{\textwidth}{X}
\hline
\textbf{Inclusion Criteria}\\
\hline
\textit{\textbf{I1:}} Papers that address misuse detection for security APIs, including papers that use either an automated or manual approach or rely on existing tools to verify the usage of security APIs.
\\
\textit{\textbf{I2:}} Papers published in peer-reviewed venues that are ranked by CORE or Scimago.\\
\textit{\textbf{I3:}} Papers written in English and their full text are accessible. \\
\hline
\textbf{Exclusion Criteria}\\
\hline
\textit{\textbf{E1:}} Papers that target detecting vulnerabilities in the internal design or implementation of security APIs, not misuses.\\
\textit{\textbf{E2:}} Papers that target detecting misuses of generic APIs. \\
\textit{\textbf{E3:}} Short papers less than six pages.\\
\textit{\textbf{E4:}} Book chapters, dissertations, and non-peer-reviewed publications (e.g., keynotes, editorials, tutorials, and panel discussions).\\
\hline
\end{tabularx}
\end{table}

\subsection{Selection of the Primary Studies}\label{subsec:study_selection}
\begin{figure}[h]
  \centering  \includegraphics[width=0.95\textwidth]{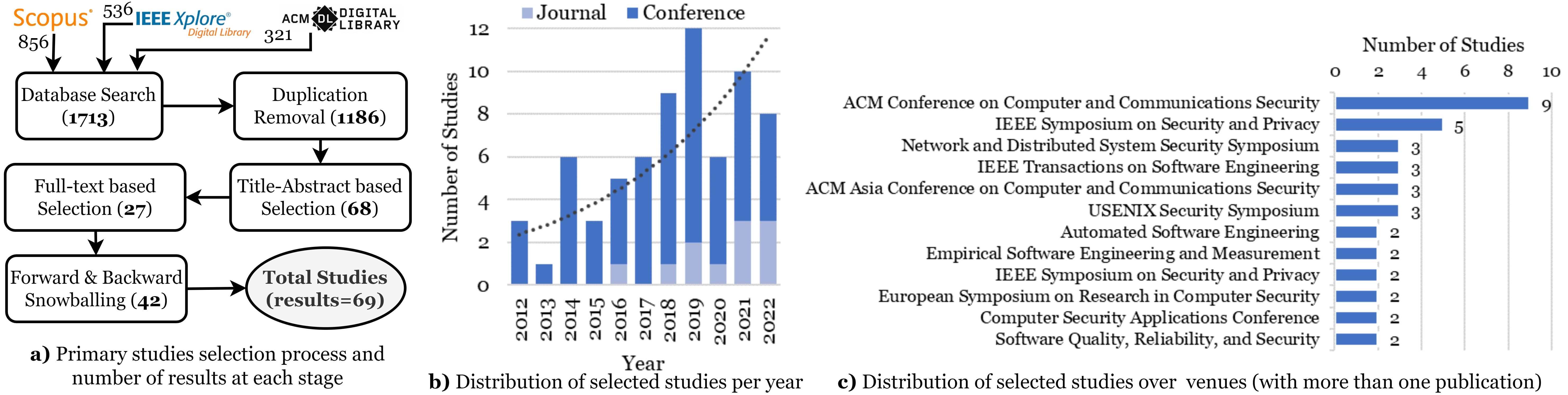}
  \caption{Primary studies selection process and their distribution over years and publication venues}
  \label{fig:paper_sel_dist}
\end{figure}

Figure~\ref{fig:paper_sel_dist}.a illustrates the different phases involved in the study selection process. 
In November 2022, we executed the search string in our data sources without any time limit on publication year, resulting in 1,713 studies. 
To remove duplicates, we initially compiled a pool of unique papers from Scopus, totaling 856. Subsequently, we supplemented this pool with additional unique papers from IEEE Xplore (271 papers) and then ACM DL (59 papers).
Next, we refined our search by applying the inclusion-exclusion criteria outlined in Table~\ref{tab:IE}. This involved initial screening based on title and abstract, resulting in the inclusion of 59 papers from Scopus, 8 from IEEE Xplore, and 1 from ACM DL. Subsequently, further refinement was performed through full-text assessment, leading to the selection of 26 papers from Scopus and 1 from IEEE Xplore.
Furthermore, we used forward and backward snowballing techniques~\cite{snowball_guidelines} to ensure the maximum number of relevant papers were included in our review. This involves examining the citations and references of the selected papers to identify any missing relevant papers. Finally, 69 papers were included in our SLR, and their details can be found in our online appendix~\cite{appendix}. Each paper in the review is assigned a unique identifier (S\#). At each stage of the selection process, we meticulously deliberated and addressed any uncertainties through discussions among all the authors to minimize the risk of selection bias.

\subsection{Data Extraction and Synthesis}\label{subsec:data_extract_synthesis}
We developed a Data Extraction Form (DEF) that comprises 14 data items essential for addressing our RQs, which we elaborated in our online appendix~\cite{appendix}. Data items (D1-D6) include demographic information such as title, author, venue, publication year, and publisher. To simplify the analysis of the extracted data relevant to our RQs, we categorized the data items into the following groups: RQ1 (D7-D8: security APIs, language), RQ2 (D9-D10: misuses, consequences), RQ3 (D11-D14: technique, modeling input data type, testing input data type, output type), and RQ4 (D15-D18: evaluation strategy, evaluation metrics, dataset, misuses reported). A pilot study was conducted on 12 papers to refine the DEF for capturing the necessary information in the most effective and summarized form.

\textbf{\textit{Data Synthesis:}} We used descriptive statistics to analyze demographic information data items, while thematic analysis was used to analyze RQ-relevant data items. To conduct thematic analysis, we followed the steps outlined in the guideline by~\citep{thematic_ana}. Firstly, we familiarized ourselves with the data by reading and examining the extracted data. Next, we generated initial codes to capture security APIs, misuses, detection techniques, and evaluation methods. Then, we searched for themes and generated potential themes for each data item by merging the corresponding initial codes based on their similarities. We reviewed the themes and mapped them iteratively to ensure all codes and themes were accurately allocated. Finally, we reviewed the synthesized results for each RQ and resolved any disagreements through regular meetings to finalize the answers to RQs.

\subsection{Distribution of Studies}\label{subsec:studies_dist}

Figure~\ref{fig:paper_sel_dist}.b depicts the distribution of the 69 primary studies included in this review across different years. No relevant studies were identified before 2012, but since then, there has been an upward trend in publications toward 2022. This increasing trend signifies a growing interest from the research community 
in security API misuse detection, highlighting the need for a systematic analysis of this evolving field. The figure also shows the number of papers based on the venue type, including journals and conferences. Notably, a substantial majority—over 84\%—of the selected papers (58 out of 69) were published in conferences. Figure~\ref{fig:paper_sel_dist}.c provides further insight into publication venues, including journals and conferences with more than one publication. Security-specific venues are well-represented, with the ACM Conference on Computer and Communications Security (CCS) and the IEEE Symposium on Security and Privacy (S\&P) emerging as the most popular venues with 9 and 5 publications, respectively. Software engineering venues are also present (e.g., Automated Software Engineering with 2 papers), demonstrating the interdisciplinary nature of this field.

\section{RQ1: Security APIs} \label{sec:APIs}

This section presents our findings regarding RQ1, which focuses on the security APIs researchers have studied for the purpose of misuse detection (\S~\ref{sec:APIs-tax}) and their misuse trend within real-world software (\S ~\ref{sec:APIs-misuse}).

\subsection{API Taxonomy} \label{sec:APIs-tax}

Our analysis revealed APIs from various contexts studied for misuse detection. Based on their primary purpose, we established a taxonomy of six API categories: \textbf{\textit{Cryptographic primitives}}, for essential low-level cryptographic functions, \textbf{\textit{SSL/TLS}}, for secure network communications, \textbf{\textit{OAuth}}, for access delegation without sharing credentials, \textbf{\textit{Fingerprint}}, for fingerprint-based authentication, \textbf{\textit{Spring Security}}, for integration of authentication and authorization mechanisms, and \textbf{\textit{SafetyNet Attestation}}, for device and application integrity checks. We further identified key security functionalities addressed by these APIs, including confidentiality, integrity (comprising data, device, and application integrity), authentication, and authorization. Table~\ref{tab:APIs} presents API categories with further details on their functionalities, supported programming languages, specific API instances, and references from the reviewed studies. As shown, cryptographic primitives (31 studies) have received the most research attention, followed by SSL/TLS (14 studies). Notably, there are 12 additional studies that investigated both APIs. OAuth has been the focus of 9 studies, while Fingerprint, Spring Security, and SafetyNet Attestation each have been explored in one study.

Regarding the programming language, the majority of research appears to be focused on Java security APIs, as indicated by Table~\ref{tab:APIs}. This focus could stem from the complex design of Java APIs [S20, S60], highlighting the demand for the development of misuse detection approaches in Java. Additionally, Java holds significant popularity among developers across various domains and platforms such as web applications, mobile apps, enterprise systems, and embedded software~\cite{PYPL}. Aside from Java, there were studies dedicated to security APIs in other programming languages, including C/C++ (8 studies), Python (3 studies), JavaScript (1 study), and Go (1 study) (Table~\ref{tab:APIs}).

\begin{table}[t]

\centering
\caption{Security APIs and their mappings with primary studies (number of primary studies indicated in parentheses)}
\label{tab:APIs}
\footnotesize
\begin{tblr}{
    width = \linewidth,
    rowsep = 0.25pt,
  colspec = {Q[110]Q[120]Q[55]Q[307]Q[345]},  
  cell{1}{1-5} = {c},
  cell{2-13}{1-3} = {c},
  cell{2}{1} = {r=5}{c},
  cell{2}{2} = {r=5}{c},
  cell{7}{1} = {r=2}{c},
  cell{7}{2} = {r=2}{c},
  hline{1, 2, 7,9, 10, 11, 12,13} = {1-5}{},
  hline{3,4,5,6,8} = {3-5}{},
}
\textbf{APIs}                         & \textbf{Functionality}                                   & \textbf{Language}   & \textbf{Instances}                                                                                              & \textbf{Study
  Refs}\\
\textbf{Cryptographic}\newline \textbf{Primitives}\newline (43)      & Confidentiality\newline Data Integrity\newline Authentication   & Java       & Java Cryptography Architecture (JCA), Java Cryptography Extension (JCE), BouncyCastle (BC), Jasypt, Keyczar, GNU Crypto, SunJCE, SpoungyCastle, LP11                                                       & S1, S2, S3, S4, S5, S7, S10, S11, S12, S15, S16, S17, S18, S20, S23, S27, S29, S31, S32, S34, S35, S36, S40, S41, S42, S43, S44, S45, S46, S55, S58, S61, S63, S64, S65, S66 \\
                             &                                                 & Python     & PyCrypto, PyNaCl, M2Crypto, cryptography.io, Keyczar,
  ucryptolib           & S20, S59, S60                                                                                                                                                     \\
                             &                                                 & C/C++      & CommonCrypto, Libsodium, Nettle, TomCrypt, LibTomCrypt, Libgcrypt, WolfCrypt                                             & S6, S13, S28                                                                                                                                                      \\
                             &                                                 & JavaScript & WebCrypto
  APIs                                                                                       & S56                                                                                                                                                                 \\
                             &                                                 & Go         & Go
  cryptographic APIs                                                                                & S62                                                                                                                                                                 \\
\textbf{SSL/TLS}\newline (26)                   & Confidentiality\newline Data Integrity\newline Authentication & Java       & Java Secured-Socket Extension (JSSE)                                                                              & S1, S2, S3, S8, S14, S19, S24, S26, S27, S31, S33, S34, S39, S43, S44, S45, S46, S57, S58, S66, S67, S68                                                                                             \\
                             &                                                 & C/C++      & OpenSSL, GnuTLS, Libcrypto, Libcrypt, Cryptlib, WolfSSL & S9, S21, S22, S25, S26, S68    \\
\textbf{OAuth}\newline (9)
   & Authentication\newline
  Authorization                 & -          & OAuth
  APIs provided by service providers such as Google or Facebook                                            & S30, S47, S48, S49, S50, S51, S52, S53, S54                                                                                                                                           \\
\textbf{Fingerprint}\newline (1)                  & Authentication\newline
  Authorization                 & Java       & Google
  Fingerprint API                                                                               & S37                                                                                                                                                                 \\
\textbf{Spring}\newline \textbf{Security} (1)                       & Authentication\newline
  Authorization                 & Java       & Spring
  framework                                                                                     & S38  \\
\textbf{SafetyNet}\newline
  \textbf{Attestation} (1)      &Device/App\newline Integrity                     & Java       & Google
  SafetyNet Attestation                                                                         & S69                  
\end{tblr}

\end{table}

The following sections provide an overview of each security API, focusing on key components and functionalities. 

\subsubsection{Cryptographic primitives APIs}

Cryptography is an essential component of secure software development as it plays a crucial role in maintaining confidentiality, data integrity, and authenticity.  Developers often utilize APIs that implement cryptography primitives (referred to as crypto APIs, hereafter) to integrate these features into their software. Cryptography primitives are fundamental building blocks of cryptography. They consist of low-level functions including \textbf{\textit{(i) symmetric encryption}}, \textbf{\textit{(ii) asymmetric encryption}}, \textbf{\textit{(iii) hash and message authentication code}}, \textbf{\textit{(iv) key derivation}}, \textbf{\textit{(v) key storage}}, and \textbf{\textit{(vi)~pseudorandom number generator}}. 

\begin{figure}[thb]
  \centering
  \includegraphics[width=0.9\textwidth]{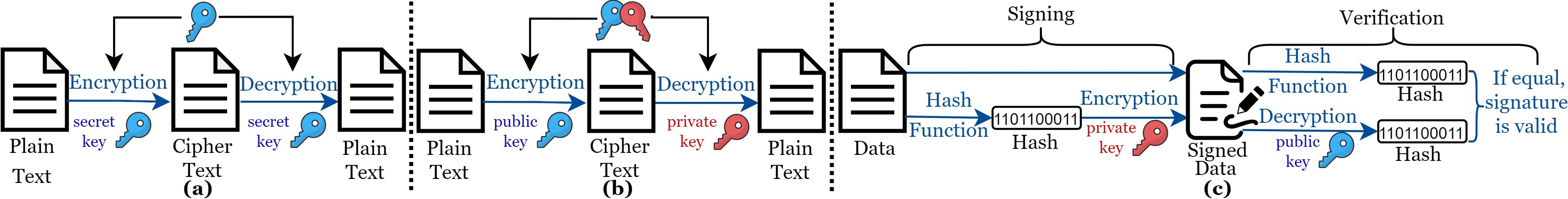}
  \caption{ a) Symmetric Encryption b) Asymmetric Encryption c) Signing and Verification in Digital Signature}
  \label{fig:crypto}  
\end{figure} 

\noindent\textit{\textbf{Symmetric Encryption:}} Symmetric encryption algorithms or \textit{ciphers}, secure data by converting \textit{plaintext} into \textit{ciphertext} that only authorized entities can decrypt. Symmetric encryption, known as private key cryptography, uses the same key for encryption and decryption, as illustrated in Figure~\ref{fig:crypto}.a. Block ciphers are the most prevalent type of symmetric encryption that divide plaintext into fixed-size blocks and encrypt them into ciphertext blocks of the same size.

\noindent\textit{\textbf{Asymmetric Encryption:}} Asymmetric encryption, also known as \textit{public-key cryptography}, uses two distinct keys, a \textit{public key} and a \textit{private key}. As shown in Figure~\ref{fig:crypto}.b, the public key is used to encrypt the data, while the private key is used to decrypt it. In addition, asymmetric encryption can be used to implement \textit{digital signatures} for ensuring authenticity in communications. To this end, the sender signs data with a private key, and the receiver verifies the signature with the sender's public key. Figure~\ref{fig:crypto}.c
demonstrates this process.

\noindent\textit{\textbf{Hash and Message Authentication Code:}} 
Hash functions maintain data integrity by converting input data of arbitrary length into unique and fixed-length hash values. As slight changes in the input result in completely different hashes, hash functions are effective for detecting any modification to the original data. Message Authentication Codes (MACs), while similar to hash functions, also incorporate a secret key. This key allows the sender to authenticate their identity as the message's origin, thereby ensuring both authenticity and integrity.

\noindent\textit{\textbf{Key Derivation:}} A Key Derivation Function (KDF) generates a cryptographic key from a \textit{password} or \textit{passphrase} that fulfills standards such as minimum length, entropy, and brute-force resistance. It is commonly employed in combination with \textit{Password-Based Encryption (PBE)}. The process of key derivation through a KDF typically involves applying a hash function, using a random value, called \textit{salt}, for an adequate number of \textit{iterations} to prevent brute-force attacks.
    
\noindent\textit{\textbf{Key Storage:}} 
Preserving the confidentiality and integrity of encrypted data in cryptography heavily relies on proper key storage practices. Key storage algorithms are designed to assist developers in securely storing sensitive credentials, such as key material. These algorithms require a strong \textit{password} or \textit{passphrase} as input to provide adequate security.

\noindent\textit{\textbf{PseudoRandom Number Generator:}} 
Randomness plays a crucial role in all aspects of cryptography.
Cryptography APIs offer PseudoRandom Number Generator (PRNG) functions to ensure the generated number holds the requisite level of randomness for cryptographic applications. PRNGs rely on a seed for generating random numbers that must also be random to prevent any potential predictability in the generated numbers.

Our survey covered 43 studies examining the usage of crypto APIs. Of these, 36 focused on the Java Cryptography Architecture (JCA), while the Python PyCrypto API and the C/C++ CommonCrypto API were each examined in 3 studies, and JavaScript and Go were each covered in 1 study.

\subsubsection{SSL/TLS APIs} 

\begin{figure}[t]
  \centering
  \includegraphics[width=0.9\textwidth]{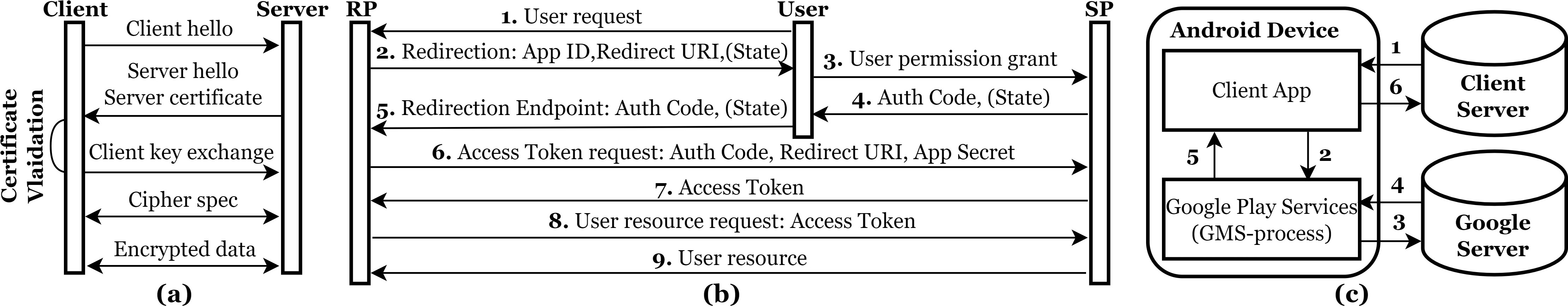}
  \caption{ (a) An overview of a simplified SSL handshake, (b) OAuth authorization code grant flow between user, relying party (RP), and service provider (SP), and (c) attestation process performed using Google SafetyNet Attestation API [S69]}
  \label{fig:ssl_oauth_attestation}
  
\end{figure}

Secure Sockets Layer (SSL) and Transport Layer Security (TLS) protocols are used to establish a secure channel for communication between a client and a server, protecting them from potential attacks like a Man-in-the-Middle (MitM). 
These protocols rely on cryptographic primitives to ensure the authentication, confidentiality, and integrity of network messages. To establish and validate SSL connections, developers can utilize SSL/TLS APIs such as OpenSSL~\cite{openssl}, which encapsulate the details and functionalities of these protocols. 
A critical aspect of SSL connection establishment is authenticating the server. During the SSL handshake, shown in Figure~\ref{fig:ssl_oauth_attestation}.a, the server presents its \textit{public key certificate} to the client as a means of authentication. It is essential that the client carefully verifies the authenticity of the server's certificate to ensure the security of the SSL connection. The validation process involves \textbf{\textit{certificate validation}} and \textbf{\textit{hostname verification}}. 

In \textbf{\textit{certificate validation}}, a client must carefully verify that the certificate has been issued and signed by a trusted \textit{Certificate Authority (CA)} and has not expired or been revoked. Additionally, the client must validate the certificate chain presented by the server. This involves verifying each certificate in the chain and ensuring that each one is issued by the CA immediately above it, with a trusted root CA at the top of the chain. 

In \textbf{\textit{hostname verification}}, a client must verify that the \textit{hostname} included in the certificate matches the hostname that the client is attempting to connect to. This verification process prevents MitM attacks, where an attacker intercepts the communication between a client and server, and impersonates the server by sending false information to the client. 

In our survey, 26 studies focused on the usage of SSL/TLS APIs, with 22 studies for Java APIs, e.g., Java Secure Socket Extension (JSSE), and 6 studies for C/C++ APIs, e.g., OpenSSL.

\subsubsection{OAuth APIs}

Open Authorization (OAuth)~\cite{hardt2012rfc} is a popular authorization protocol that allows end-users to grant third-party websites or applications access to their private resources on a remote server without sharing their credentials.
This process involves three major roles: the \textbf{\textit{user}} or \textbf{\textit{resource owner}} who owns protected resources, the \textbf{\textit{Service Provider (SP)}} that hosts the resources, and \textbf{\textit{Relying Party (RP)}}, known as \textit{client application} that uses the SP to obtain access to the user’s resources. \textit{\textbf{Access tokens}} are issued by SPs to RPs with the owner's approval for accessing protected resources.
OAuth APIs provided by SPs such as Google, Twitter, or Facebook are used by developers to authenticate users or obtain access to users’ resources through their major accounts in SPs.
Although OAuth was first introduced as an authorization framework, it has been widely adopted to implement Single-Sign-On authentication, making it difficult for developers to use it properly. To address this challenge, OpenID Connect~\cite{sakimura4openid} was introduced as an authentication framework based on OAuth. 
The OAuth specification~\cite{hardt2012rfc} defines four different protocol flows or grant types: \textbf{\textit{(i)~authorization code}}, \textbf{\textit{(ii) implicit}}, \textbf{\textit{(iii) resource owner password credentials}}, and \textbf{\textit{(iv) client credentials}}. 
Figure~\ref{fig:ssl_oauth_attestation}.b depicts the process of authorization code grant, which is the most commonly used grant type.
As shown in Figure~\ref{fig:ssl_oauth_attestation}.b, the process begins with a user sending a request to an RP to access a remote resource (step 1). The RP then redirects the user to the SP with an \textit{APP ID} and an optional \textit{state} parameter to bind this request (step 2). Next, the user authenticates with the SP and grants the RP's requested permissions (step 3). The SP issues an \textit{authorization code} and an optional \textit{state} parameter to the user (step 4).
The user is then redirected back to the RP's redirection endpoint, where the request is rejected if the received \textit{state} parameter mismatches the initial one (step 5).
Next, the RP  sends the \textit{authorization code} and its \textit{secret} (established during registration with the SP) to the SP to request an access token (step 6).
The SP verifies the RP app by validating the \textit{App ID} and app \textit{secret} and then responds with an \textit{access token} (steps 7). 
With the access token, the RP requests user data from the SP, which is then shared with the RP accordingly (steps 8-9). 
In the implicit grant (which is simpler than the authorization code grant), in step 4, the SP directly responds with an access token instead of an authorization code, without authenticating the RP. Resource owner password credentials and client credentials grants are rarely used. 

Our review included 8 studies 
that evaluated the usage of OAuth APIs in Android and web applications. Additionally, one study~[S54] focused on both OAuth and OpenID Connect APIs for implementing authentication in Android apps.

\subsubsection{Fingerprint APIs}

Our review identified one study  [S37] that investigated the usage of Fingerprint API in Android apps.
Both Google~\cite{Google_guidline} and OWASP~\cite{OWASP_guidline} guidelines recommend using a fingerprint reader in conjunction with cryptographic operations for secure authentication. This involves using the fingerprint to unlock a cryptographic key protected by the \textit{Trusted Execution Environments (TEE)}, rather than just recognizing the user. TEE, an integral part of modern smartphones, can securely generate and store cryptographic keys. Combining TEE with fingerprint readers for Two-factor authentication provides strong security comparable to external hardware devices such as YubiKeys~\cite{YubiKeys}.
To interact with the fingerprint sensor and verify whether a legitimate user has touched it, four essential steps are required to follow [S37]. 
That begins with 
\textbf{\textit{generating a cryptographic key}} where developers specify key properties via parameters such as setting the \textit{user authentication required} parameter to \textit{True}, ensuring key usability only after a legitimate user has touched the fingerprint reader. 
Next, 
\textbf{\textit{the key is unlocked through user authentication}}. If a legitimate user touches the sensor, the cryptographic key is unlocked, triggering a series of callback functions. Developers can 
\textbf{\textit{override the fingerprint callbacks}} to handle different scenarios based on user legitimacy. 
Once authenticated, \textbf{\textit{the unlocked key can be used}} by an app to encrypt, decrypt, or sign data. Google recommends using a previously generated private key to sign a server-provided authentication token to authenticate, and then to send this token to the app's remote backend~\cite{Google_guidline}.

\subsubsection{Spring Security APIs}
Spring Security~\cite{spring-security} is a powerful and highly customizable framework for securing Java-based applications. It provides a wide range of security services, such as authentication, authorization, and access control. While Spring Security provides its own set of authentication features, it also supports integration with various authentication mechanisms such as \textit{Lightweight Directory Access Protocol (LDAP)}, \textit{OAuth}, and \textit{Java Database Connectivity (JDBC)}. The framework enables developers to implement role-based and permission-based authorization, allowing for granular access control policies for different parts of their applications.
It allows the implementation of an access-control specification model by typically defining various filters for finding access to a given resource.
In addition, it offers features for securing communication between different application components using HTTPS, SSL/TLS, and other encryption mechanisms. 
In our review, one study [S38] examined Spring Security's use for implementing an access-control specification model.

\subsubsection{SafetyNet Attestation APIs}
Attackers can alter an application's behavior either by directly modifying the app or by obtaining root access to the host system and injecting malicious code. Hence, developers need to ensure their app's code integrity and the client device's status.
Google offers the SafetyNet Attestation API~\cite{Google-att} to check the integrity of a device or application and detect compromised devices and tampered applications. 
Figure~\ref{fig:ssl_oauth_attestation}.c illustrates the attestation process using this API. 
The \textit{attest} function, requiring a \textit{nonce} and an \textit{API Key}, triggers the attestation API. The nonce is generated by the application's backend server and sent to the device upon attestation request (step 1). The API Key is created using Google's \textit{API Console}, a platform for developers to manage their Google APIs. When attestation is requested (step 2), \textit{Google Mobile Services (GMS)} conducts several checks on the device, forwarding the results to Google’s server (step 3). 
Google’s server responds with signed attestation data (step 4), which GMS delivers to the client (step 5). The client app extracts a \textit{JSON Web Signature (JWS)} from the data, sending it to the backend server for validation (step 6), followed by the client server verifying the JWS (step 7). In our review, one study [S69] analyzed the usage of the attestation API in Android applications.

\subsection{Misuse Trends} \label{sec:APIs-misuse}

A total of 50 primary studies have investigated real-world software to identify security API misuse. This section provides insights into their findings (\S~\ref{sec:API-misuse-trend}) and analyzes the variations observed (\S~\ref{sec:API-misuse-desc}).

\subsubsection{Security API misuse}\label{sec:API-misuse-trend}
\begin{figure}
\begin{minipage}[c]{0.4\textwidth}
\includegraphics[width=\linewidth]{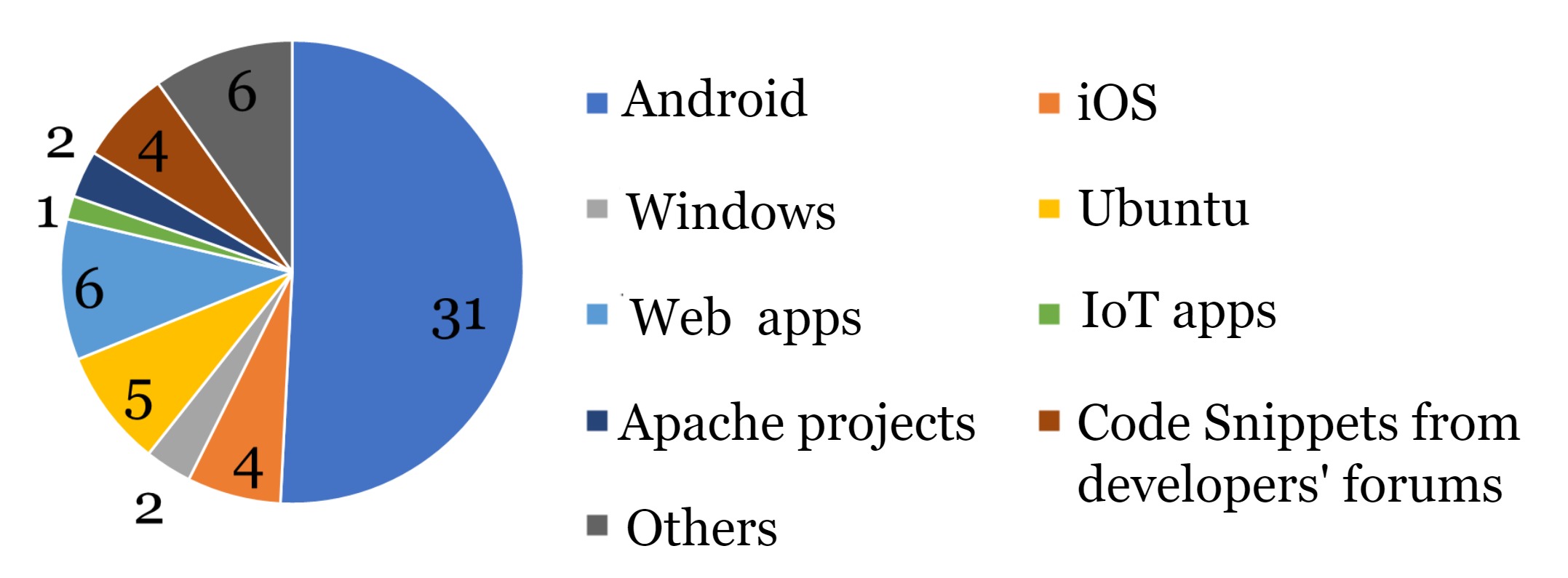}
\caption{Distribution of types of software artifacts analyzed by reviewed studies}
\label{fig:dist-soft}
\end{minipage}
\hfill
\begin{minipage}[c]{0.5\textwidth}
\includegraphics[width=\linewidth]{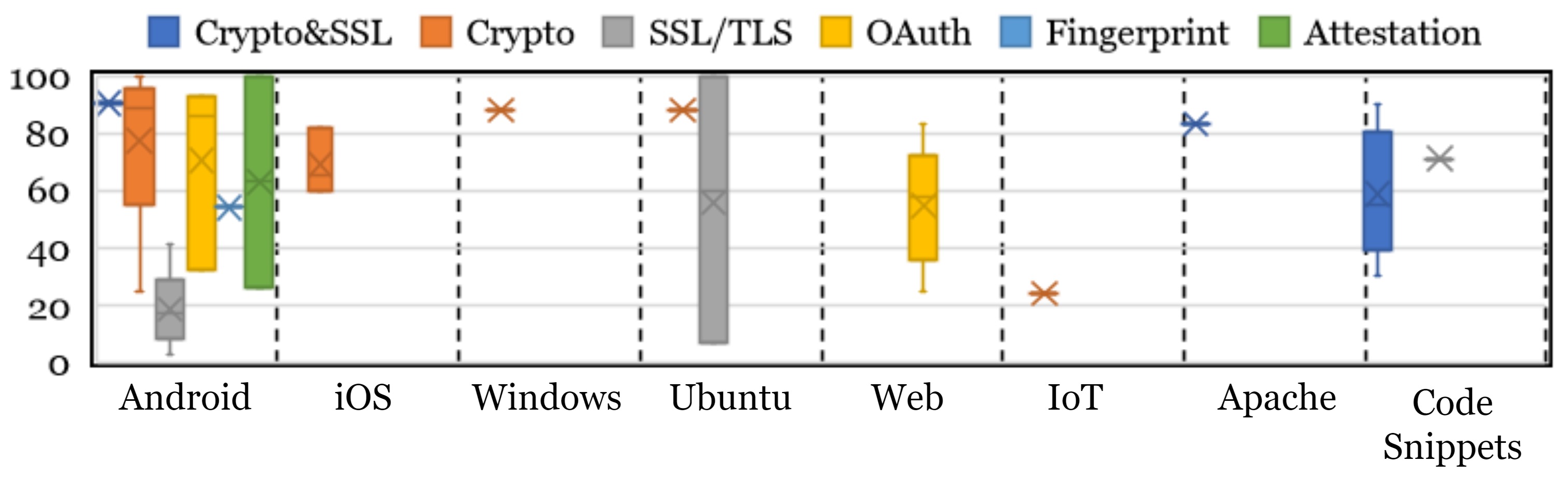}
\caption{Boxplots with mean markers illustrating the percentage of software artifacts with at least one misuse (misuse rate)}
\label{fig:API-results}
\end{minipage}

\end{figure}

Figure~\ref{fig:dist-soft} demonstrates diverse types of software analyzed and their distribution over reviewed studies. Given the widespread use of mobile devices for storing confidential data and Android being the most widely used OS, Android apps have received the most significant attention from researchers studying security API misuse. Many studies used the Google Play Store as the primary source for obtaining Android apps. Several studies also used AndroZoo~\cite{andozoo2016}, which offers the most extensive publicly available dataset of Android apps collected from the Google Play Store. Researchers also analyzed iOS apps from the official Apple App Store. Additionally, two studies leveraged Maven Central and GitHub to analyze Apache projects, and one study [S22] assessed 521 IoT firmware obtained from various IoT vendors' websites. There were also studies dedicated to assessing Windows and Ubuntu applications, and studies that investigated OAuth API usage in web applications. In addition, four studies examined the state of security API usage in code snippets found in developers' forum posts. This holds significance as prior research revealed that forums like StackOverflow (SO) are crucial sources of information for developers, who may use code snippets from such sources in their software projects~\cite{fischer2019stack}.  Three studies assessed 187 [S33], 3,834 [S45], and 25,855 [S66] code snippets from SO, while another study [S64] investigated 140, 71, and 48 posts from Oracle Java Cryptography (OJC), Google Android Developers (GAD), and Google Android Security Discussions (GASD), respectively.

The findings from these studies reveal a concerning trend of security API misuse among developers, with some studies reporting misuses in nearly 100\% of the applications examined. For example, one study [S11] analyzed ten thousand Android apps and found that 95\% of them contained at least one crypto misuse. Similarly, another study [S45] exploring a dataset of Android applications discovered that 15\% of these apps incorporated security-related code snippets from Stack Overflow, of which 98\% were found to contain misuses.
Figure~\ref{fig:API-results} summarizes the findings, illustrating misuse rates for each API across various software types. Misuse rate, as defined here, represents the percentage of software artifacts identified with at least one security API misuse.

\subsubsection{Variations in findings}\label{sec:API-misuse-desc}

As shown in Figure~\ref{fig:API-results}, there are wide ranges of misuse rates for some APIs. For instance, crypto misuse rates in Android applications range dramatically from 25\% to 100\%. Even code snippets from developer forums exhibit a wide range of misuse rates (48\%-90\%). Similar variations are observed for other APIs and platforms, such as SSL misuse rates in Ubuntu, which vary significantly from 7\% to 100\%. There are various reasons for these substantial differences.
We identified five key factors contributing to these variations, including \textbf{\textit{dataset size, data source, temporal consideration, misuse scope}}, and \textbf{\textit{detection techniques}}.

First, the size of the datasets analyzed in each study can vary considerably. Android studies, for example, encompass investigations targeting a mere 45 applications [S15] to massive analyses exceeding 500,000 apps [S36].
Second, the data for analysis has been collected from various sources. For instance, the study exploring developer forums, [S64], reported differing misuse rates—90\% in OJC, 71\% in GAD, and 48\% in GASD, reflecting the diverse nature of these platforms. It is important to consider that GASD is not exclusively focused on programming, potentially contributing to the lower misuse rate observed.
Third, studies were conducted across different years and utilized datasets collected at various time intervals. Security practices evolve over time, potentially affecting misuse prevalence. For example, a study [S12] compared datasets from 2012 and 2016 to understand changes in misuse prevalence. Their analysis revealed a decrease in specific misuse types, likely due to increased developer awareness of insecure practices over time. Conversely, new misuses might be discovered over time, leading to apparent increases in misuse rates.
Fourth, studies also differ in their focus and scope of misuse. Some delve into a single, specific misuse, while others aim for a broader analysis. Additionally, the prevalence of different misuse types naturally varies. The specific types and prevalence of misuses identified will be discussed in detail in Section~\ref{sec:Misuses}.
Last but not least, the detection techniques employed in these studies can introduce variations in the reported misuse rates. Some techniques may have limitations, such as reporting false positives or non-exploitable misuses. For example, a study [S69] initially reported a 100\% misuse rate of SafetyNet Attestation in Android apps integrating this API, but after conducting exploit attempts, it found that only a small set of apps had exploitable misuses. Section~\ref{sec:tech} will dive deeper into the details of detection techniques.

Despite the varied findings, it is evident that security API misuse remains a significant challenge for developers, and our study represents a timely effort to raise awareness about this critical aspect of secure software development and to provide fundamental knowledge for the development of effective detection approaches.

\section{RQ2: Misuses of Security APIs} \label{sec:Misuses}

\begin{figure}[th]
    \begin{minipage}[c]{0.27\textwidth}
        \includegraphics[width=\linewidth]{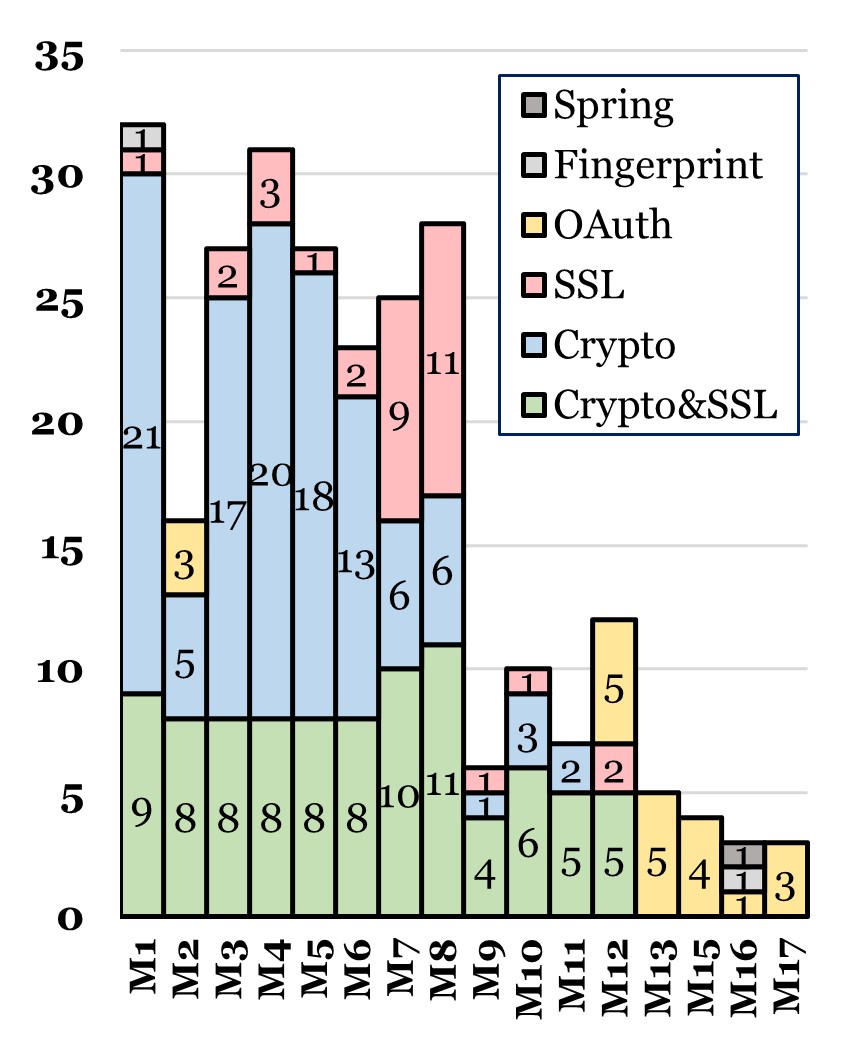}
        \caption{Number of studies for misuses grouped by APIs (including only misuses with multiple studies)}
        \label{fig:misuse-dist}
    \end{minipage}
    \hfill
    \begin{minipage}[c]{0.72\textwidth}
        \includegraphics[width=\linewidth]{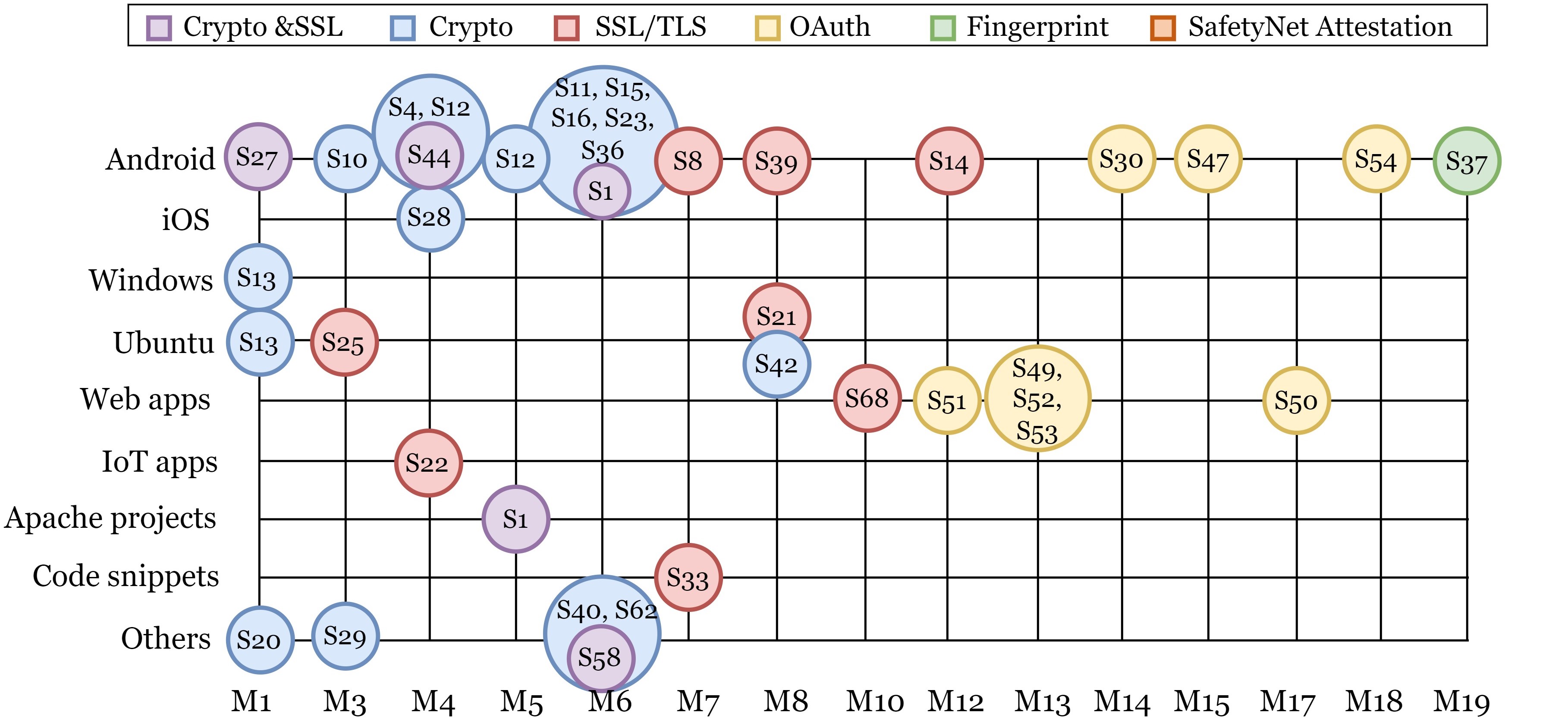}
        \caption{Most common security API misuses in real-world software\\ identified by primary studies}
        \label{fig:most_common_misuses}
    \end{minipage}
    \
\end{figure}

Through thorough analysis and semantic categorization, we identified 30 distinct misuse types, each assigned a unique identifier (\textit{M}\#). Table~\ref{tab:misuse_tax} lists these misuses and their mappings to relevant security APIs, while Figure~\ref{fig:misuse-dist} illustrates the number of research efforts dedicated to each misuse type. As shown, \textit{insecure cryptographic key management} (\textit{M1}), \textit{insecure cryptography algorithms} (\textit{M4}), and \textit{improper SSL certificate validation} (\textit{M8}) are the most extensively explored misuses in research. Notably, they are also among the most common misuses reported for crypto and SSL APIs. Figure~\ref{fig:most_common_misuses} illustrates different findings regarding the most common misuses of security APIs identified in real-world software by reviewed studies. \textit{Insecure cryptography algorithms (broken hash)} for crypto API, \textit{improper certificate validation} for SSL/TLS API, \textit{lack or misuse of the \textit{State} parameter (M13)} for OAuth API, and \textit{lack or misuse of cryptography (M19)} for Fingerprint API are the misuses seen with the highest frequency in this figure. In the following, we elaborate on these misuse types, discussing their consequences and prevalence in software applications.

\begin{table}[]
    \centering
    \caption{Taxonomy of security API misuses and mappings with security APIs (C for Cryptography, T for SSL/TLS, O for OAuth, F for Fingerprint, S for Spring Security, and A for SafetyNet Attestation)}
    \label{tab:misuse_tax}
    {\footnotesize
    \setlength{\tabcolsep}{2pt} 
    \renewcommand{\arraystretch}{1} 
    \begin{tabularx}{.95\linewidth}{XccccccXcccccc}
        \hline
        \textbf{Misuse} & \textbf{C} & \textbf{T} & \textbf{O} & \textbf{F} & \textbf{S} & \textbf{A} & \textbf{Misuse} & \textbf{C} & \textbf{T} & \textbf{O} & \textbf{F} & \textbf{S} & \textbf{A} \\
        \hline
        \textbf{M1)} Insecure cryptographic key management & \checkmark & \checkmark & - & \checkmark & - & - &
        \textbf{M16)} Lack or misuse of authentication & - & - & \checkmark & \checkmark & \checkmark & -\\
        \textbf{M2)} Insecure storage of credentials & \checkmark & \checkmark & \checkmark & - & - & - &
        \textbf{M17)} Insecure OAuth grant types & - & - & \checkmark & - & - & -\\
        \textbf{M3)} Insecure cryptographic PRNGs\textsuperscript{1} & \checkmark & \checkmark & - & - & - & - &
        \textbf{M18)} Lack of PKCE parameters in OAuth & - & - & \checkmark & - & - & -\\
        \textbf{M4)} Insecure configurations for encryption & \checkmark & \checkmark & - & - & - & - &
        \textbf{M19)} Lack or misuse of cryptography & - & - & - & \checkmark & - & -\\
        \textbf{M5)} Insecure configurations for PBE\textsuperscript{2} & \checkmark & \checkmark & - & - & - & - &
        \textbf{M20)} Lack of authorization & - & - & - & - & \checkmark & -\\
        \textbf{M6)} Insecure cryptography algorithms & \checkmark & \checkmark & - & - & - & - &
        \textbf{M21)} Incorrect authorization & - & - & - & - & \checkmark & -\\
        \textbf{M7)} Improper SSL hostname verification & \checkmark & \checkmark & - & - & - & - &
        \textbf{M22)} Spring method call with higher access rights & - & - & - & - & \checkmark & -\\
        \textbf{M8)} Improper SSL certificate validation & \checkmark & \checkmark & - & - & - & - &
        \textbf{M23)} Local attestation checks & - & - & - & - & - & \checkmark\\
        \textbf{M9)} Improper SSL socket & \checkmark & \checkmark & - & - & - & - &
        \textbf{M24)} Sending incomplete data to attestation server & - & - & - & - & - & \checkmark\\
        \textbf{M10)} Insecure SSL/TLS standard & \checkmark & \checkmark & - & - & - & - &
        \textbf{M25)} Local nonce generation for attestation & - & - & - & - & - & \checkmark\\
        \textbf{M11)} Improper error handling & \checkmark & \checkmark & - & - & - & - &
        \textbf{M26)} Verification flaws in attestation & - & - & - & - & - & \checkmark\\
        \textbf{M12)} Lack of SSL protection & \checkmark & \checkmark & \checkmark & - & - & - &
        \textbf{M27)} Using test server for attestation & - & - & - & - & - & \checkmark\\
        \textbf{M13)} Insecure \textit{state} management in OAuth & - & - & \checkmark & - & - & - &
        \textbf{M28)} Null or wrong API key & - & - & - & - & - & \checkmark\\
        \textbf{M14)} OAuth client-side API call & - & - & \checkmark & - & - & - &
        \textbf{M29)} Using deprecated API & - & - & - & - & - & \checkmark\\
        \textbf{M15)} Insecure OAuth redirection options & - & - & \checkmark & - & - & - &
        \textbf{M30)} Performing attestation only at first launch & - & - & - & - & - & \checkmark\\
        \hline 
    \end{tabularx}
    }
    
    \begin{flushleft}
    \footnotesize
    1. Pseudorandom Number Generators, 2. Password-Based Encryption
    \end{flushleft} 
    
\end{table}

\noindent\textit{\textbf{M1—Insecure cryptographic key management:}}
M1 is the most extensively studied misuse in the literature, specifically on \textit{\textbf{crypto}}, \textit{\textbf{SSL/TLS}}, and \textbf{\textit{Fingerprint}} APIs (Figure~\ref{fig:misuse-dist}). It also ranks among the most common misuses found in various software applications across Android, Windows, and Ubuntu (Figure~\ref{fig:most_common_misuses}). In the context of crypto and SSL/TLS, M1 often arises from practices like \textit{deterministic key generation} and \textit{poor key derivation}, leading to predictable and easily compromised keys. The use of \textit{hard-coded, static (constant), expired, previously-used} or any kind of \textit{predictable} keys is among the misuses identified by prior research.
Listing~\ref{lst:m1m4} illustrates an example of hard-coded keys identified within an Android app. Keys of \textit{inadequate length} also increase the possibility of a brute-force attack; for example, RSA keys and ECC keys should be at least 2048 and 224 bits long, respectively~\cite{barker2009recommendation}. 
Additionally, the practice of \textit{retaining keys in memory} after their intended use creates opportunities for code injection or side-channel attacks, enabling potential key recovery [S13]. Another significant concern arises from \textit{insecure key distribution}. Using a key agreement protocol that allows a single peer to generate the shared secret without involving the other peers compromises the security of the negotiated key [S13]. Likewise, \textit{key exchange without authentication} from a trusted entity leads to vulnerabilities to MitM attacks, where malicious servers can impersonate trusted servers and gain access to sensitive information [S15]. Furthermore, \textit{ignoring integrity checks} is a notable concern. Developers should conduct integrity checks after symmetric key exchange to ensure the integrity of the exchanged keys [S29].
In the context of \textbf{\textit{Fingerprint}}, M1 often arises from a \textit{lack of cryptographic key generation} when developers neglect using methods for generating keys to be used for securing fingerprint-based authentication [S38]. Another concern is the \textit{use of unlocked keys}, where the key remains unlocked, allowing attackers with root access to use it without requiring the user to touch the fingerprint sensor [S38].

\begin{multicols}{2}
\begin{lstlisting} [language=java, 
                   label={lst:m1m4}, 
                   abovecaptionskip=0pt, 
                   escapeinside=@@, 
                   lineskip=-0.5ex, 
                   caption={\footnotesize \fboxsep=0pt\colorbox{lightcoral}{M1} and \fboxsep=0pt\colorbox{lightcyan}{M4} identified in an Android app \texttt{[S19]}.}]
    public class AESUtil {
      String @\fboxsep=0pt\colorbox{lightcyan}{iv = "0102030405060708";}@
      String @\fboxsep=0pt\colorbox{lightcoral}{key = "czabcd1234czabcd";}@ 
    
      String Encrypt(String sSrc) throws Exception {
        if (key == null || key.length() != 16) {
          return null;}
    
        byte[] @\fboxsep=0pt\colorbox{lightcoral}{raw = key.getBytes("UTF-8");}@ 
        SecretKeySpec @\fboxsep=0pt\colorbox{lightcoral}{s = new SecretKeySpec(raw, "AES");}@ 
        Cipher ci = Cipher.getInstance("AES/CBC/Iso10126Padding");
        IvParameterSpec ps;
        @\fboxsep=0pt\colorbox{lightcyan}{ps = new IvParameterSpec(iv.getBytes("UTF-8"));}@
        ci.init(Cipher.ENCRYPT_MODE, s, ps);
        byte[] encrypted = ci.doFinal(sSrc.getBytes("UTF-8"));
        return new String(Base64.encodeBase64(encrypted),"UTF-8");}}
\end{lstlisting}
\begin{lstlisting} [language=java, 
                   label={lst:m2}, 
                   abovecaptionskip=0pt, 
                   escapeinside=@@, 
                   lineskip=-0.5ex, 
                   caption={\footnotesize \fboxsep=0pt\colorbox{lightcoral}{M2} within a popular application-server artefact  \texttt{[S11]}.}]
    protected String getKeystorePassword() {
        String keyPass = (String)attributes.get("keypass");
        if (keyPass == null) { @\fboxsep=0pt\colorbox{lightcoral}{keyPass = "changeit";}@}
        String keystorePass = (String)attributes.get("keystorePass");
        if (keystorePass == null) { keystorePass = keyPass;}
        return keystorePass;}
\end{lstlisting}

\begin{lstlisting} [language=java, 
                   label={lst:m3}, 
                   abovecaptionskip=0pt, 
                   escapeinside=@@, 
                   lineskip=-0.5ex, 
                   caption={\footnotesize \fboxsep=0pt\colorbox{lightcoral}{M3} identified in an Android app \texttt{[S19]}.}]
    void onCreate(Bundle savedInstanceState) {
        super.onCreate(savedInstanceState);
        setContentView(R.layout.activity_main);    
        SecretKeySpec secretKeySpec = null;
        SecureRandom @\fboxsep=0pt\colorbox{lightcoral}{sr = SecureRandom.getInstance("SHA1PRNG")}@;
        @\fboxsep=0pt\colorbox{lightcoral}{sr.setSeed("any data used as random seed"}@.getBytes()); ...}
\end{lstlisting}
\end{multicols}


\noindent\textit{\textbf{M2—Insecure storage of credentials:}} 
Several studies examining \textbf{\textit{crypto}}, \textbf{\textit{SSL/TLS}}, and  \textbf{\textit{OAuth}} APIs highlight a concerning trend: developers often compromise security by using insecure storage practices for credentials (passwords, secret keys, etc.).
Storing credentials in plain text \textit{strings} leaves them vulnerable to credential dumping as programming languages like Java do not clear string values from memory unless the garbage collector runs~\cite{Oracle2022JDK}.
Similarly, storing sensitive data in \textit{files} or \textit{shared storage} poses a significant risk of data exposure. It is recommended to use the secure \texttt{\small{key storage}} method offered by crypto APIs, which requires a password for access~\cite{barker2016nist}. However, developers often undermine its security by using \textit{hard-coded, static (constant), or predictable passwords} (Listing~\ref{lst:m2}). In the context of OAuth, M2 arises from the \textit{local storage of RP secrets and access tokens} [S30, S48]. Attackers can reverse engineer client-side applications to steal locally stored RP secrets and impersonate legitimate applications to request access tokens from SPs. Stolen access tokens can also grant unauthorized access to user accounts and sensitive information stored on SPs.

\noindent\textit{\textbf{M3—Insecure PseudoRandom Number Generators (PRNGs): }} Insecure \texttt{\small{PRNGs}} are a major source of \textbf{\textit{crypto}} and \textbf{\textit{SSL/TLS}} vulnerabilities~\cite{bernstein2013factoring}. It is essential to exclusively use the secure \texttt{\small{PRNG}} offered by crypto APIs and use a truly randomly generated seed for initialization. However, developers often make two common mistakes: 1) using \textit{simple \texttt{\small{PRNGs}}} that are known to be insecure as they can generate predictable random numbers~\cite{krawczyk1992predict}, and  2) using \textit{static (constant), low-entropy, predictable or previously-used} seeds, as depicted in Listings~\ref{lst:m3} and~\ref{lst:m3m6}. These practices severely undermine the security of cryptographic materials, like keys, that rely on randomness for their strength.

\noindent\textit{\textbf{M4—Insecure configurations for encryption:}} One common misuse of \textbf{\textit{crypto}} and \textbf{\textit{SSL/TLS}} APIs is using \textit{unsafe modes of operation} for encryption, such as \texttt{\small{Electronic Codebook (ECB)}}. \texttt{\small{ECB}} mode encrypts data blocks independently, transforming identical message blocks into identical ciphertext blocks, thus revealing data patterns and compromising confidentiality. To ensure security, it is recommended to use more secure modes, such as \texttt{\small{Cipher Block Chaining (CBC)}} or \texttt{\small{Galois/Counter Mode (GCM)}}. Other instances of unsafe encryption modes include \texttt{\small{DESede}} with \texttt{\small{ECB}}, \texttt{\small{DES}} with \texttt{\small{CBC3 SHA}}, \texttt{\small{AES}} with \texttt{\small{CBC}} and \texttt{\small{PKCS5Padding}}, \texttt{\small{CBC}} without \texttt{\small{HMAC}}, and \texttt{\small{3DES}} with \texttt{\small{EDE CBC SHA}}.
Additionally, \textit{Initialization Vectors (IVs)} are crucial in several encryption modes to add entropy to ciphertexts. To ensure the security of cryptographic schemes, IVs must be randomly and properly generated. However, some developers introduce vulnerabilities by using \textit{empty, zeroed, hard-coded, static, badly-derived (e.g., deriving from keys or messages), short-length, previously-used, or any kind of predictable IVs}. Listing~\ref{lst:m1m4} illustrates an example of hard-coded IVs identified within an Android app.
Another parameter that requires secure configuration is the \textit{padding} scheme, which specifies how to fill the last block of data in encryption. Missing padding or using insecure padding (e.g., \texttt{\small{PKCS 1-v1.5}} for \texttt{\small{RSA}}) makes it easier for an attacker to launch a padding oracle attack and recover the plaintext.

\noindent\textit{\textbf{M5—Insecure configurations for Password-Based Encryption (PBE):}} Within \textbf{\textit{crypto}} and \textbf{\textit{SSL/TLS}} APIs, \texttt{\small{PBE}} relies on carefully chosen parameters including \textit{salt}, \textit{password}, and \textit{iteration count} to derive strong encryption keys. Improperly setting these parameters can significantly compromise the security of the derived key. 
Common misconfigurations of \texttt{\small{PBE}} include using an \textit{empty, static (constant), short-length (size<64 bits~\cite{kaliski2017rfc}), or predictable salt} or using a \textit{hard-coded, static, weak, NIST-blacklisted, expired, previously-used, or predictable password}, which introduces vulnerabilities to brute-force and dictionary attacks. 
Moreover, developers may prefer to choose \textit{small iteration counts} (less than 1000~\cite{kaliski2017rfc}) to achieve better performance, making it easier for attackers to perform brute-force attacks. Listing~\ref{lst:m5} showcases M5 present in an Android application. The code uses an iteration count based on the password length, falling far below the recommended minimum. Additionally, the salt is generated by hashing the password itself, which introduces another security risk. An attacker gaining access to the salt could potentially recover the password [S1].

\noindent\textit{\textbf{M6—Insecure cryptography algorithms:}} 
Algorithms once considered secure can become vulnerable due to newly discovered weaknesses and attacks. This makes it challenging for developers to keep up with the latest updates. Thus, one of the most common misuses of \textbf{\textit{crypto}} and \textbf{\textit{SSL/TLS}} APIs involves using outdated algorithms including: \textit{ unsafe symmetric encryption algorithms such as 64-bit block ciphers (e.g., \texttt{\small{DES}}, \texttt{\small{IDEA}}, \texttt{\small{Blowfish}}, \texttt{\small{RC4}}, \texttt{\small{RC2}}), weak PBE algorithms (e.g., \texttt{\small{PBKDF1}}), insecure asymmetric ciphers (e.g., \texttt{\small{RSA}}, \texttt{\small{ECC}}), insecure cryptographic MACs and broken hash functions (e.g., \texttt{\small{SHA1}}, \texttt{\small{MD5}}, \texttt{\small{MD4}}, \texttt{\small{MD2}}) as well as insecure combinations of encryption and hashes or MACs (e.g., \texttt{\small{PBKDF}} with \texttt{\small{SHA224}})}. Listing~\ref{lst:m3m6} shows the use of insecure M5 for hashing found in a code snippet from developers' forums.

\begin{multicols}{2}
\begin{lstlisting} [language=java, 
                   label={lst:m5}, 
                   abovecaptionskip=0pt, 
                   escapeinside=@@, 
                   lineskip=-0.5ex, 
                   caption={\footnotesize \fboxsep=0pt\colorbox{lightcoral}{M5} identified in an Apache project \texttt{[S1]}.}]
PBEKeySpec getPBEParameterSpec(String password) throws Throwable {
    MessageDigest md = MessageDigest.getInstance(MD_ALGO); // MD5
    byte[] @\fboxsep=0pt\colorbox{lightcoral}{saltGen = md.digest(password.getBytes())}@;
    byte[] salt = new byte[SALT_SIZE];
    System.arraycopy(saltGen, 0, salt, 0, SALT_SIZE);
    int @\fboxsep=0pt\colorbox{lightcoral}{iteration = password.toCharArray().length + 1;}@
    return new PBEKeySpec(password.toCharArray(), salt, iteration);}
\end{lstlisting}
\begin{lstlisting} [language=java, 
                   label={lst:m3m6}, 
                   abovecaptionskip=0pt, 
                   escapeinside=@@, 
                   lineskip=-0.5ex, 
                   caption={\footnotesize \fboxsep=0pt\colorbox{lightcoral}{M3} and \fboxsep=0pt\colorbox{lightcyan}{M6} in a code snippet from developers' forums  \texttt{[S64]}.}]
    // weak hash of user's password
    @\fboxsep=0pt\colorbox{lightcyan}{md = MessageDigest.getInstance("MD5");}@
    byte[] hash = md.digest(password.getBytes());
    // weak PRNG with fixed seed
    @\fboxsep=0pt\colorbox{lightcoral}{sr = SecureRandom.getInstance("SHA1PRNG");}@
    @\fboxsep=0pt\colorbox{lightcoral}{sr.setSeed(hash.getBytes());}@
    byte[] keyBytes = new byte[}
\end{lstlisting}
\end{multicols}

\noindent\textit{\textbf{M7—Improper SSL hostname verification:}} Hostname verification is a crucial security measure that ensures the hostname in the SSL certificate matches the server hostname to which the client is trying to connect. However, various studies in the context of \textbf{\textit{crypto}} and \textbf{\textit{SSL/TLS}} APIs [e.g., S1, S2, S8, S9] have revealed that some developers \textit{trust all hostnames} or do not verify the hostname correctly. For instance, the code snippet in Listing~\ref{lst:m7}, sourced from SO, always returns true, thereby accepting all hostnames. Improper hostname verification enables an attacker to intercept the communication between the client and the server by presenting a valid SSL certificate for a poisoned hostname. 

\noindent\textit{\textbf{M8—
Improper SSL  certificate validation:}} 
Many developers make mistakes in implementing proper certificate validation as identified by several studies in the context of \textbf{\textit{crypto}} and \textbf{\textit{SSL/TLS}} APIs [e.g., S14, S15, S19]. One of the most common mistakes is blindly \textit{trusting all certificates}, allowing attackers to present fake certificates and gain unauthorized access to sensitive information. An example is shown in Listing~\ref{lst:m8}, where validation is implemented through empty methods. Additionally, some developers only check that each certificate in the chain has not expired without performing any other validation. 
Other ways of compromising certificate validation include \textit{incomplete validation, neglecting to check for expiration or revocation, trusting self-signed certificates, trusting too many CAs, trusting certificates with unclear names, inadequate CA verification, or insecure certificate pinning}.


\begin{multicols}{2}
\begin{lstlisting} [language=java, 
                   label={lst:m7}, 
                   abovecaptionskip=0pt, 
                   escapeinside=@@, 
                   lineskip=-0.5ex, 
                   caption={\footnotesize \fboxsep=0pt\colorbox{lightcoral}{M7} in a code snippet from StackOverflow \texttt{[S45]}.}]
    //Empty HostnameVerifier - Accepts all hostnames
    public boolean verify(String hostname, SSLSession session) {
        @\fboxsep=0pt\colorbox{lightcoral}{return true;}@
    }
\end{lstlisting}
\begin{lstlisting} [language=java, 
                   label={lst:m8}, 
                   abovecaptionskip=0pt, 
                   escapeinside=@@, 
                   lineskip=-0.5ex, 
                   caption={\footnotesize \fboxsep=0pt\colorbox{lightcoral}{M8} identified in an Android app \texttt{[S43]}.}]
class r$b implements X509TrustManager {
    public void checkClientTrusted(X509Certificate[], String) @\fboxsep=0pt\colorbox{lightcoral}{\{\}}@
    public void checkServerTrusted(X509Certificate[], String) @\fboxsep=0pt\colorbox{lightcoral}{\{\}}@
    public X509Certificate[] getAcceptedIssuers() { return null;}}   
\end{lstlisting}
\end{multicols}

\noindent\textit{\textbf{M9—Improper SSL socket:}} In the context of \textbf{\textit{SSL/TLS}} API, the SSL socket is aimed to establish a connection between a specific host and a specific port. Nonetheless, verifying and authenticating the server's hostname is essential before establishing the connection. A flawed implementation of the SSL socket may ignore hostname verification when creating the socket, as depicted in Listing~\ref{lst:m9} [S1].

\noindent\textit{\textbf{M10—Insecure SSL/TLS standard:}} TLS, the successor of SSL, is generally considered to be more secure. However, older versions of TLS, including TLS 1.0 and TLS 1.1, have been found to be susceptible to various types of attacks, such as POODLE, BEAST, and CRIME, and therefore are no longer deemed secure. These outdated versions have been deprecated~\cite{barnes2015deprecating, k2021deprecating, turner2011prohibiting}, and TLS 1.2 is being recommended as the minimum protocol version for secure communication. Nevertheless, several studies on \textbf{\textit{crypto}} and \textbf{\textit{SSL/TLS}} APIs have revealed that some developers still use outdated versions of TLS and compromise the security of transmitted data [S2, S66, S68, etc].


\noindent\textit{\textbf{M11—Improper error handling:}} Some developers prioritize functionality over security, leading them to disregard errors, as identified in use cases of the \textbf{\textit{crypto}}, \textbf{\textit{SSL/TLS}}, and \textbf{\textit{SafetyNet Attestation}} APIs. For crypto and SSL/TLS, developers may ignore errors occurring during certificate validation and simply proceed with normal operations [S24, S34, S43, etc.]. Listing~\ref{lst:m11} exemplifies this by using an empty method for exception handling in an Android app. Likewise, errors may occur during the integrity checks performed by the SafetyNet Attestation API, and disregarding them can result in the failure of the attestation process [S69].



\begin{multicols}{2}
\begin{lstlisting} [language=java, 
                   label={lst:m9}, 
                   abovecaptionskip=0pt, 
                   escapeinside=@@, 
                   lineskip=-0.5ex, 
                   caption={\footnotesize M9 identified in an Android app \texttt{[S1]}.}]
    try {   SSLContext instance = SSLContext.getInstance("TLS");
            // ... 
            this.webSocketClient.setSocket(instance.getSocketFactory().createSocket());
    } catch (Throwable e) { ... }
    this.webSocketClient.connect();
\end{lstlisting}
\begin{lstlisting} [language=java, 
                   label={lst:m11}, 
                   abovecaptionskip=0pt, 
                   escapeinside=@@, 
                   lineskip=-0.5ex, 
                   caption={\footnotesize \fboxsep=0pt\colorbox{lightcoral}{M11} identified in an Android app \texttt{[S1]}.}]
void checkServerTrusted(X509Certificate[] chain, String str){
    try {
        this.f7427a.checkServerTrusted(chain, str);
    } 
    catch (CertificateException e) @\fboxsep=0pt\colorbox{lightcoral}{\{\}}@ //Ignores exception   
}
\end{lstlisting}
\end{multicols}


\noindent\textit{\textbf{M12—Lack of SSL protection:}} M12 represents a prevalent misuse identified for \textbf{\textit{crypto}}, \textbf{\textit{SSL/TLS}}, and \textbf{\textit{OAuth}} APIs. While using crypto and SSL/TLS, \textit{occasional use of HTTP} exposes the application to potential attacks like SSL stripping~\cite{marlinspike2009more, marlinspike2009new}, wherein a malicious actor can launch a MitM attack on an SSL connection [e.g., S1, S5, S9, S10].  For OAuth, security heavily relies on the secure transmission of messages throughout the process. \textit{Transmitting messages in plaintext without SSL/TLS encryption} enables attackers to eavesdrop and pilfer access tokens or other OAuth credentials [S30, S47, S50, S51]. Listing~\ref{lst:m12} shows the use of a raw access token for OAuth transmissions in an Android app.

\begin{multicols}{2}

\begin{lstlisting} [language=java, 
                   label={lst:m12}, 
                   abovecaptionskip=0pt, 
                   escapeinside=@@, 
                   lineskip=-0.5ex, 
                   caption={\footnotesize \fboxsep=0pt\colorbox{lightcoral}{M12} (raw OAuth access token) in an Android app \texttt{[S30]}.}]
    String aToken = getAccessToken();
    HttpClient httpClient = new DefaultHttpClient();
    HttpPost httpPost = new HttpPost("/backend.com/tokensignin");
    List<NameValuePair> params = new ArrayList<>(1);
    params.add(new BasicNameValuePair("access_token", @\fboxsep=0pt\colorbox{lightcoral}{aToken}@));
    httpPost.setEntity(new UrlEncodedFormEntity(params));    
    httpClient.execute(httpPost);
\end{lstlisting}

\begin{lstlisting} [language=javascript, 
                   label={lst:m13}, 
                   abovecaptionskip=0pt, 
                   escapeinside=@@, 
                   lineskip=-0.5ex, 
                   caption={\footnotesize M13 (no state parameter) in example code for connecting to Microsoft’s Live Connect Services \texttt{[S52]}.}]
    this.getAuthUrl = function () {
        var scopes = ['wl.signin', 'wl.basic', 'wl.offline_access', 'office.onenote_create'];
        var query = toQueryString({
            'client_id': clientId,
            'scope': scopes.join(' '),
            'redirect_uri': redirectUrl,
            'display': 'page',
            'locale': 'en',
            'response_type': 'code'  });
        return oauthAuthorizeUrl + "?" + query;};
\end{lstlisting}

\begin{lstlisting} [language=java, 
                   label={lst:m14}, 
                   abovecaptionskip=0pt, 
                   escapeinside=@@, 
                   lineskip=-0.5ex, 
                   caption={\footnotesize \fboxsep=0pt\colorbox{lightcoral}{M14} identified in an Android app \texttt{[S30]}.}]
String url = "api.provider.com/..";
HttpURLConnection c=(HttpURLConnection) new URL(url).openConnection();
c.setRequestMethod("POST");
// ...
List<String> params = new ArrayList<>();
params.add("oauth_consumer_key=" + client_id);
params.add("oauth_token=" + access_token);
params.add("oauth_signature=" + getSignature(client_secret));
// ...
c.setRequestProperty("Authorization", createHeaders(params));
String user_id = parseJSON(c.getInputStream(), "id");
@\fboxsep=0pt\colorbox{lightcoral}{newUserLogin(user\_id);}@ 
\end{lstlisting}

\begin{lstlisting} [language=java, 
                   label={lst:m15}, 
                   abovecaptionskip=0pt, 
                   escapeinside=@@, 
                   lineskip=-0.5ex, 
                   caption={\footnotesize \fboxsep=0pt\colorbox{lightcoral}{M15} identified in an Android app \texttt{[S30]}.}]
    WebView webview;
    String url = "provider.com/..?client_id=".."&redirect_url=".."&response=code";
    ...
    public void onCreate() {
        @\fboxsep=0pt\colorbox{lightcoral}{webview.loadUrl(url);}@
    }
\end{lstlisting}
\end{multicols}

\noindent\textit{\textbf{M13—Insecure state management in OAuth:}} 
The \texttt{\small{state}} parameter protects user sessions in \textbf{\textit{OAuth}} transactions against Cross-Site Request Forgery (CSRF) attacks by verifying request authenticity. In CSRF, an attacker uses a user's previous session data to make a malicious request on their behalf [S52]. OAuth guidelines recommend generating and validating a randomized \texttt{\small{state}} parameter, bound to the user's session to prevent such attacks~\cite{hardt2012rfc}.
However, developers may misunderstand its purpose leading to mistakes such as \textit{using a constant or predictable value, enabling multiple replays, neglecting \texttt{\small{state}} parameter verification, accepting requests without a \texttt{\small{state}} parameter, or assuming that all \texttt{\small{state}} parameters generated by their app are valid without proper session binding checking} [S51]. Listing~\ref{lst:m13} illustrates an example where OAuth is implemented without using a \texttt{\small{state}} parameter. This example is provided by Microsoft to help developers use their API. Nonetheless, it can mislead developers who copy the example without understanding the security implications.

\noindent\textit{\textbf{M14—OAuth client-side API call:}} 
A significant security concern in \textbf{\textit{OAuth}} authentication flows arises from the reliance on client-side API calls, which attackers could easily manipulate [S30]. Listing~\ref{lst:m14} presents a code snippet from an Android app wherein an access token is exchanged for a user ID via an API call from the app to authenticate users.

\noindent\textit{\textbf{M15—Insecure OAuth redirection options:}} To ensure the security of \textbf{\textit{OAuth}} transactions, it is crucial to use secure methodologies for handling redirection~\cite{hardt2012rfc}. 
Insecure redirection methods can allow attackers to redirect users to arbitrary domains or URLs, potentially leading to further attacks or data theft.
For instance, in a mobile context, using WebView is considered insecure as it undermines the isolation between an SP and an RP [S30]. A malicious RP can use the WebView of their mobile applications to host an SP, allowing them to access the user's cookies and log in on the user's behalf. Listing~\ref{lst:m15} exemplifies M15 with the use of WebView for redirection in an Android application.

\noindent\textit{\textbf{M16—Lack or misuse of authentication:}} 
This misuse has been identified in use cases of \textbf{\textit{OAuth}}, \textbf{\textit{Fingerprint}}, and \textbf{\textit{Spring Security}} APIs. In OAuth transactions, an SP is responsible for authenticating an RP, and vice versa.
However, a study [S47] on a collection of Android apps revealed that none of the RP apps in their investigation verified the SP's identity. Furthermore, M16 was observed in the context of the Fingerprint API, where developers designate authentication methods as null [S37]. Additionally, research by [S38] found M16 in Spring Security usage, where developers neglect to implement authentication for accessing a resource. This can occur if a developer forgets to include an authentication filter or improperly configures a filter, allowing unrestricted access to the resource (CWE-306)~\cite{CWE-306}. 

\noindent\textit{\textbf{M17—Insecure OAuth grant types:}}
The security of \textbf{\textit{OAuth}} transactions highly depends on the choice of grant type. It is essential to avoid using insecure grant types, such as implicit for authentication. Implicit grants raise a major security concern because the access token is not bound to the intended RP, which enables an attacker to use a user's access token, issued to the malicious application, to log in as the user on a benign application [S48]. Best current practices recommend using the authorization code flow, which can be protected by PKCE [S54].

\noindent\textit{\textbf{M18—Lack of PKCE parameters for OAuth authorization code grant:}} The authorization code grant is generally considered to be the most secure grant type for \textbf{\textit{OAuth}}. However, it remains susceptible to code interception attacks, where an attacker intercepts the authorization code sent by the SP and uses it to obtain an access token~\cite{rahat2022cerberus}. To mitigate this vulnerability, the Proof Key for Code Exchange (PKCE) protocol was introduced in the OAuth 2.0 specification~\cite{sakimura2015proof}. PKCE verifies that the requesting application is the same one that originally requested the authorization code by using a cryptographically linked code verifier and code challenge exchanged between the application and the SP. PKCE is recommended as a mandatory security measure for public clients to secure the authorization code grant.


%


%


\noindent\textit{\textbf{M19—Lack or misuse of cryptography in fingerprint authentication:}} This misuse occurs while using \textbf{\textit{Fingerprint}} API when developers do not utilize any cryptography operation after the user touches the sensor or perform an insecure cryptography operation using constant encryption keys [S37].

\noindent\textit{\textbf{M20—Lack of authorization:}}  While using \textbf{\textit{Spring Security}}, a developer may fail to include appropriate authorization filters for a particular resource, which needs valid authorization according to the access-control specification model. Using authentication as the authorization filter also results in the same misuse since it only verifies the user's identity. Both scenarios lead to the vulnerability known as missing authorization (CWE-862)~\cite{CWE-862}, allowing unrestricted access to resources, either for all users or just authenticated users, based on the type of applied filter [S38].

\noindent\textit{\textbf{M21—Incorrect authorization:}}
This misuse arises from an incorrect authorization formula while using an authorization filter in \textbf{\textit{Spring Security}} [S38]. The misuse, known as the vulnerability of incorrect authorization (CWE-863)~\cite{CWE-863}, results in unauthorized users gaining access to resources.

    
\noindent\textit{\textbf{M22—Spring method call with higher access rights:}} 
M22 is another instance of the incorrect authorization vulnerability (CWE-863). It arises when a developer properly configures a resource while using \textbf{\textit{Spring Security}}, yet calls a method requiring higher access rights in a deeper application layer, which should not be accessible to the user [S38]. 

%


\noindent\textit{\textbf{M23—Local attestation checks:}} 
The \textbf{\textit{SafetyNet Attestation}} API returns a JWS object representing the device and application state. It is crucial to send the JWS object to the backend server for verification. Performing local checks enables an attacker to bypass the verification by modifying the application [S69]. 

\noindent\textit{\textbf{M24—Sending incomplete data to attestation server:}} The \textbf{\textit{SafetyNet Attestation}} JWS should be sent to the server for verification. However, a developer may choose to send only certain values extracted from the JWS object. This enables attackers to replace the missing values on a compromised device or application without any means for servers to detect tampering [S69].

\noindent\textit{\textbf{M25—Local nonce generation for attestation:}} 
\textbf{\textit{SafetyNet Attestation}} relies on nonces used in the \texttt{\small attest} function to prevent replay attacks. Nonces are included in the JWS output and checked against the value passed to the function to confirm that the correct JWS result is being attested. However, if the nonce value is generated locally on a compromised device or application, an attacker can exploit a previously generated nonce value to conduct a replay attack [S69]. 

\noindent\textit{\textbf{M26—Verification flaws in attestation:}} The verification process of \textbf{\textit{SafetyNet Attestation}} JWS involves several checks by the server, such as validating the nonce, APK package name, and the hash of the application's signing certificates present in the JWS payload. Inaccurate or incomplete execution of these validations may enable an attacker to send a tampered SafetyNet JWS to the server and bypass the verification [S69].

\noindent\textit{\textbf{M27—Using test server for attestation:}} 
Google provides a verification service for \textbf{\textit{SafetyNet Attestation}}, which is essentially a test server that allows a client application to submit a JWS for verification.
It is important to note that this service is exclusively designed for testing purposes, and using it in a production environment may compromise the security of SafetyNet Attestation [S69].

\noindent\textit{\textbf{M28—Null or wrong API key:}} Developers must provide the \textbf{\textit{SafetyNet Attestation}} API with a valid key obtained from the Google APIs Console. However, it is not uncommon for developers to mistakenly use an incorrect or null API key, leading to an error in the attestation process [S69]. If this error is not handled properly, the attestation process fails, leaving any tampering undetected.

\noindent\textit{\textbf{M29—Using deprecated API:}} The attestation process cannot be accomplished if developers use the deprecated \textbf{\textit{SafetyNet Attestation}} API, which always returns an error and can not generate a valid SafetyNet JWS [S69].

\noindent\textit{\textbf{M30—Performing attestation only at first launch:}} \textbf{\textit{SafetyNet Attestation}} should be consistently performed during an application life cycle, specifically when launching or handling sensitive information. However, some developers only perform SafetyNet Attestation during the first launch, leaving the application vulnerable to tampering [S69]. An attacker could exploit this by initially launching the application in a non-tampered state, then subsequently tampering with the device or application without detection, as SafetyNet Attestation would no longer be performed.
\color{black}
\section{RQ3: Misuse Detection Techniques} \label{sec:tech}
\begin{figure}[t]
  \centering
  \includegraphics[width=\textwidth]{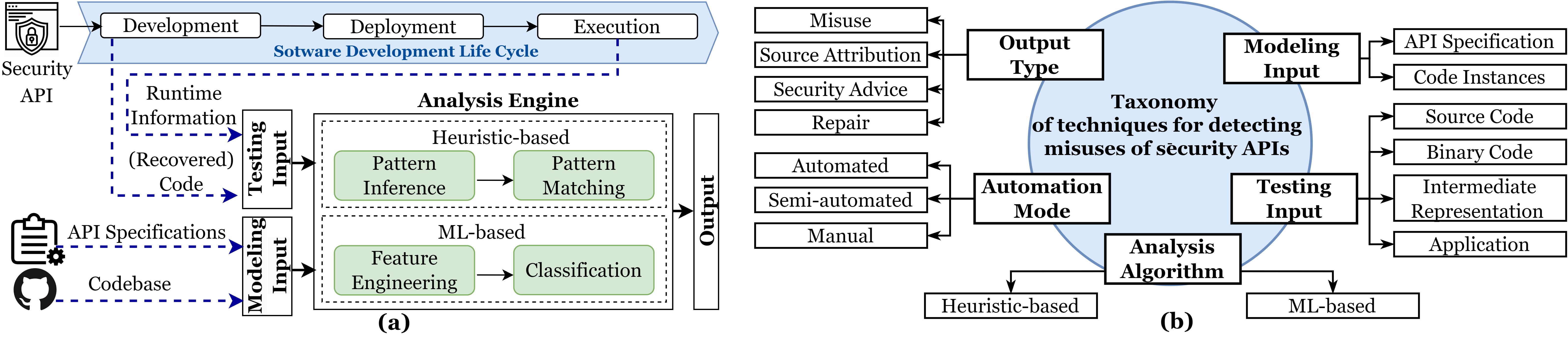}
  \caption{(a) Overview and (b) taxonomy of techniques used by primary studies for detecting misuses of security APIs}
  \label{fig:overview_taxonomy}
  
\end{figure}

Figure~\ref{fig:overview_taxonomy}.a introduces a high-level taxonomy for misuse detection techniques employed by reviewed studies. It identifies four key components: \textbf{\textit{Modeling Input}}, the data used to build the detection model (including API specifications or code examples), \textbf{\textit{Testing Input}}, the data used to test the software for identifying misuses (including code or runtime information), \textbf{\textit{Analysis Engine}}, the core for identifying misuses that can be implemented manually, semi-automatically, or fully automated using heuristics- or Machine Learning (ML)-based algorithms, and \textbf{\textit{Output}}. Building upon this, Figure~\ref{fig:overview_taxonomy}.b categorizes existing literature based on these factors: Modeling Input, Testing Input, Output Type, Automation Mode, and Analysis Algorithm. The following subsections will explore each factor in detail.

\subsection{Types of Modeling Input}

\begin{figure}[h]
  \centering
  \includegraphics[width=\textwidth]{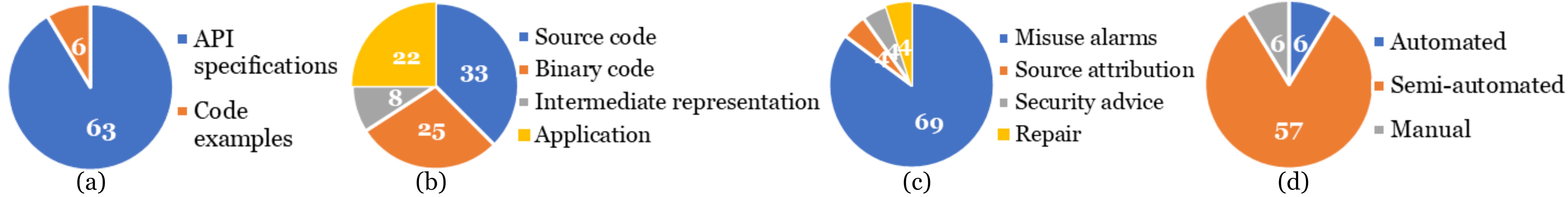}
  \caption{Distribution of misuse detection techniques adopted by primary studies over (a) modeling input types, (b) testing input types, (c) output types, (d) automation mode
}
  \label{fig:dist-techs}
  
\end{figure}

Misuse detection models are built using either \textbf{\textit{API specifications}} (63 studies) or \textbf{\textit{code examples}} (3 studies) (Figure~\ref{fig:dist-techs}.a). The most commonly used source is API specifications, which can be found in API documentation, relevant literature on security API misuse, or established standards from organizations like NIST and IETF [S10]. API specification-based detection involves manually defining misuse or normal patterns for security APIs and subsequently applying a pattern-matching technique to identify misuses. However, a key challenge lies in the evolving nature of API specifications. For instance, the SHA-1 hash function, once considered secure, is no longer recommended due to discovered vulnerabilities~\cite{stevens2017yarik}. Manual patterns defined using outdated specifications might miss new or evolving threats.

To address this challenge, recent research has explored code examples as an alternative source for inferring usage patterns. Given that code repositories are regularly updated to integrate security fixes, CryptoChecker introduced DiffCode to infer crypto misuse patterns from code changes [S29]. This involves three steps: (i) collecting code changes from GitHub repositories, focusing on patches for classes that use the target API; (ii) creating Directed Acyclic Graphs (DAGs) to represent the invoked APIs and related parameter values, and comparing them to extract relevant changes; and (iii) clustering similar changes to identify API misuse patterns. On the other hand, Seader [S2] and VuRLE [S46] leveraged Abstract Syntax Trees (ASTs) of a given vulnerable code and its corresponding repaired code to infer misuse patterns from code examples. Additionally, source code from repositories was used to train ML-based detection models to classify API use within an application as either normal or misuse [S18, S45, S65].

\subsection{Types of Testing Input}

Various input types are utilized to analyze an application for the presence of misuse, including \textit{\textbf{source code, binary code, intermediate representation,}} or \textit{\textbf{applications}}, depending on the analysis algorithm employed (Section~\ref{sec:algo} details analysis algorithms).
The most common approach is analyzing the source code, which was employed in 33 studies. Some misuse detection tools are developed as IDE plugins to provide real-time feedback as developers write code. For example, FixDroid~[S5] was introduced as a plugin for Android Studio, or CogniCrypt~[S3] as an Eclipse plugin. However, source code analysis has limitations. Existing approaches are often limited to specific programming languages. A study found developers frequently use C for cryptography in MicroPython projects, highlighting the need for multi-language tools [S20]. Additionally, source code analysis is typically restricted to open-source applications. Some studies~[S14, S23, S26] decompiled binary code to gain access to source code when the source code is unavailable. However, obfuscation techniques that modify class files to protect the source code present challenges to successful decompilation.

In the absence of source code, 25 studies employed binary code analysis.
Binary files are typically packaged with other resources, such as images and configuration files, into application files. Thus, misuse detection techniques typically require \textit{reverse engineering} to decode an APK file and disassemble it into binary files, ready for analysis. Disassembling is typically faster than decompiling and is not affected by obfuscation. Common program analysis frameworks used by primary studies to extract binary files from application files include SOOT~\cite{lam2011soot}, APKTool~\cite{Apktool}, and Androguard~\cite{Androguard}. 

However, analyzing low-level representations of applications (i.e., binary files) can be laborious and error-prone. Several studies [S4, S11, S12, S28, S37, S44] addressed this by converting binary files into a higher-level Intermediate Representation (IR) that accurately captures the program behavior.
Using IR also helps address the inconsistency of different input formats. For instance, CryptoREX~[S22] converts diverse binary files into a unified IR to carry out large-scale crypto misuse detection for IoT firmware with diverse underlying architectures. 
Another study~[S45] detected misuse by identifying instances of code snippets with security API misuse in Android apps. To achieve this, both known vulnerable code and Android apps were transformed to a unique IR using WALA~\cite{WALA}—a program analysis framework.

Meanwhile, 22 studies [e.g., S6, S8, S10, S13, S14] detected misuse by executing applications and analyzing their runtime, eliminating the need for accessing source or binary code. However, this method is resource-intensive and time-consuming due to the deployment and execution phases. Figure~\ref{fig:dist-techs}.b illustrates the distribution of testing input types across the reviewed studies. Notably, several studies employed multiple input types for analysis, which were categorized under all relevant types. For instance, some studies [e.g., S19, S24] examined potential misuses by analyzing binary code and then detected exploitable misuses at runtime by executing the applications. 

\subsection{Output Types}

Common ways to support developers in addressing misuses are issuing \textit{\textbf{misuse alarms}} (69 studies), performing \textit{\textbf{source attribution}} (4 studies), generating \textit{\textbf{security advice}} (4 studies), and assisting with \textit{\textbf{fixing}} (4 studies) (Figure~\ref{fig:dist-techs}.c).
In the following, we elaborate on output types other than misuse alarms, which were exclusively explored in the context of crypto and SSL/TLS APIs by the literature.

\noindent\textbf{\textit{Source attribution:}} 
The objective of source attribution is to determine whether a misuse originates from an application code or a third-party library.  
It is beneficial for developers to identify libraries with misuses and avoid using them, and for researchers to avoid over-counting misuses by identifying those that stem from libraries [S12]. Three studies investigated the primary source of crypto and SSL/TLS misuse and, interestingly, showed that third-party libraries are the major reason for misuse in Android applications [S1, S12] and popular Python projects in GitHub [S20]. 
Another study [S35] evaluated a large database of third-party libraries and found that crypto misuses are very common among widely used advertising libraries. It further identified affected Android apps through third-party library detection.

\noindent\textbf{\textit{Security advice:}} Developers often face challenges while trying to fix identified issues, leading to new mistakes~\cite {tupsamudre2020fixing}.  
Integrating security advice with crypto APIs has been shown to significantly improve the accuracy and security of code [S59]. Tools like FixDroid [S5] offer real-time feedback and suggestions for quick fixes within the development environment (e.g., Android Studio). Additionally, studies have explored educational approaches to help students learn the proper use of crypto APIs [S63, S17].

\noindent\textbf{\textit{Repair:}}
Studies exploring security advice and fix suggestions often lack the ability to provide customized fixes for a given vulnerable program. Only 4 studies [S2, S16, S17, S46] proposed methods for generating such fixes for crypto and SSL/TLS misuses. One simple approach involves manually crafting patch templates, as performed by CDRep [S16] and CryptoTutor [S17]. 
However, these templates have limited coverage for diverse misuse variations.
In contrast, Seader [S2] and VuRLE [S46] automatically generated customized fixes from code examples. They compared ASTs of a given vulnerable code and its corresponding repaired code to extract the edit operations necessary to transform a vulnerable program into a secure one. Using edit operations, Seader generated an abstract fix for a vulnerable program, including abstract variables that were replaced by concrete variables to customize fixes. On the other hand, VuRLE customized the edit patterns for a specific vulnerable program and applied the customized changes to repair vulnerabilities.

\subsection{Automation Mode}

The analysis techniques reviewed can be categorized by their automation level: \textbf{\textit{manual}} (6 studies), \textbf{\textit{semi-automated}} (57 studies), or \textbf{\textit{automated}} (6 studies) (Figure~\ref{fig:dist-techs}.d).  Most studies fall under the semi-automated approach, relying on predefined patterns for correct and incorrect API usage (details in Section~\ref{sec:algo}). Automated approaches learn detection models from code examples, eliminating manual effort but requiring labeled datasets for training.  Manual analysis involves human inspection of real-world code (e.g., GitHub projects [S61] or forum posts [S64, S66]) to identify misuses.


%



\subsection{Analysis Algorithms}\label{sec:algo}
We classified the reviewed analysis algorithms into two main categories: \textbf{\textit{heuristic-based}} (66 studies ) and \textbf{\textit{ML-based}} (3 studies ). This section explores the methodologies within each group.

\subsubsection{Heuristic-based}

\begin{figure}[h]
  \centering
\includegraphics[width=.95\textwidth]{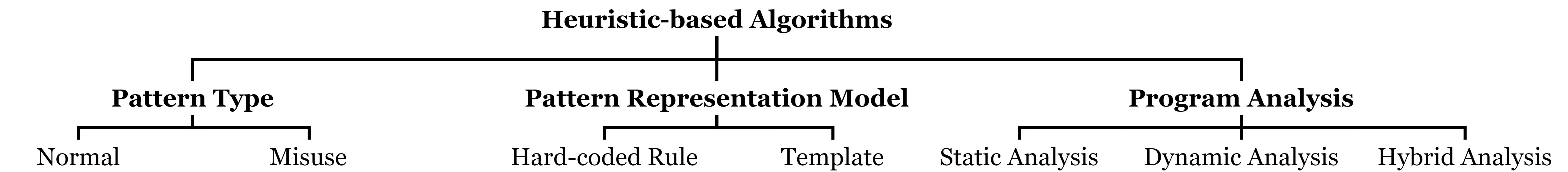}
  \caption{Taxonomy of Heuristic-based algorithms for detecting misuse of security APIs}
  \label{fig:tax-heuristic} 
\end{figure}

These algorithms typically involve inferring patterns that represent correct or incorrect usage of security APIs and then applying program analysis techniques to identify whether the application being tested matches these patterns. Figure~\ref{fig:tax-heuristic} shows the taxonomy of heuristic-based approaches based on the \textit{adopted pattern types, pattern representation models}, and \textit{program analysis techniques}, which are elaborated on below.

\noindent\textbf{A. Pattern Type:}
API misuses can be detected using two types of patterns: \textbf{\textit{normal}} (12 studies) or \textbf{\textit{misuse}} (54 studies). Misuse detection through normal patterns involves modeling correct API use and identifying any deviations from it as misuse. While APIs can be misused in numerous ways, only a small subset corresponds to proper usage. Thus, this approach can identify a wide range of misuse types through limited normal patterns. However, this approach suffers from producing high false alarms when it fails to model normal patterns thoroughly. An alternative solution is to model incorrect API uses and identify any matches with them as misuse. However, predicting all possible ways that a developer could misuse an API is a challenging task, so this approach may not capture all misuses. Nevertheless, most of the studies in our review rely on misuse patterns to avoid high false alarms in approaches based on normal patterns.

\begin{figure}[t]
  \centering
  \includegraphics[width=\textwidth]{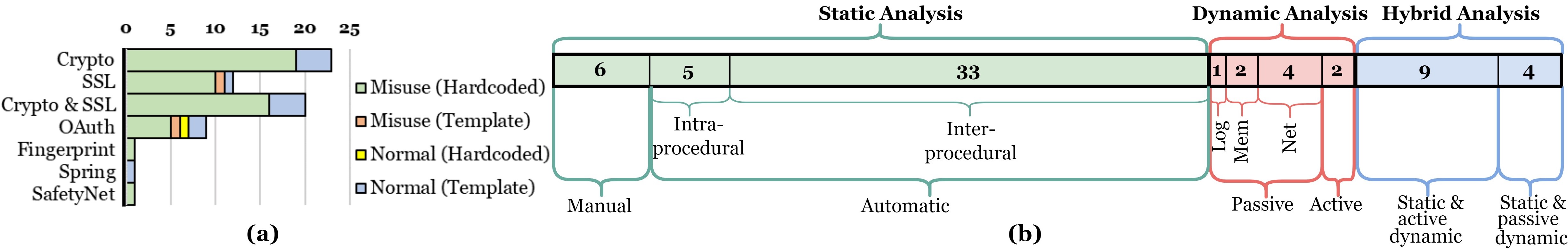}
  \caption{a) Distribution of pattern types and representation models used by primary studies for each security API b) Number of primary studies over program analysis technique in heuristic-based approaches for misuse detection}
  \label{fig:dist-heuristic}
  
\end{figure}

\noindent\textbf{B. Pattern Representation Model:} \label{sec:patt-rep}
Two main approaches for modeling patterns are \textbf{\textit{hard-coded rules}} and \textbf{\textit{templates}}. The former involves hard-coding a fixed set of rules into the misuse detection algorithm, against which applications are evaluated. For instance, \textit{``Don't use a constant key for encryption''} or \textit{``Don't store access tokens on clients''} are example rules used to detect crypto and OAuth misuses, respectively. Most studies relied on hard-coded misuse patterns to detect misuse of crypto (31 studies), SSL/TLS (21 studies), OAuth (5 studies), Fingerprint (1 study), and SafetyNet Attestation (1 study) APIs (Figure~\ref{fig:dist-heuristic}.a). CryptoLint~[S4], for instance, hard-coded 6 misuse patterns for detecting crypto misuses, which were later used and expanded in other studies [e.g., S12, S16, S29, S35]. Furthermore, one study [S54] used hard-coded normal patterns to investigate the compliance of Android apps with the best current OAuth practices.

Approaches based on hard-coded rules are straightforward to design and implement but have limitations. Notably, they are restricted to detecting only a limited set of misuses, making it difficult to extend beyond predefined rules. The ever-evolving threat landscape of security APIs requires developing more adaptable methods for pattern representation models. With this goal, some researchers proposed using templates to abstractly represent secure or insecure use of security APIs that include \textit{\textbf{language-based, graph-based, finite state machine}}, and \textit{\textbf{code-based}} templates. 


\noindent\textbf{\textit{(i) Language-based templates}} rely on a syntax-based representation of patterns. CrySL [S11], for instance, is a language designed for crypto experts to specify the normal use of crypto APIs.
Several studies [S3, S11, S36, S40, S58, S61] used CrySL to detect crypto misuses. Meta-CrySL [S55] is an extension of CrySL that helps manage variations in the API and security standards specified within CrySL. Another study [S56] introduced a formal model for security annotations that describe properties ensuring the secure usage of the WebCrypto API. Furthermore, an anti-protocol language was introduced to describe common misuse patterns for the OAuth API  [S30]. 

\noindent\textbf{\textit{ (ii) Graph-based templates}} represent key elements used while interacting with APIs as nodes and the correlations between these elements as edges. SSLint [S9], for instance, models the proper use of the SSL/TLS API based on the program dependency graph representing critical API call sites, variables, parameters, and conditions.

\noindent\textbf{\textit{(iii) Finite State Machine (FSM) templates}} represent an application's behavior while using an API through a finite number of states and transitions between them. For example, two studies [S51, S53] modeled the regular operation of OAuth using FSMs, where sending an HTTP(S) request or receiving an HTTP(S) response triggers the transition between states. FSMs were also used to model misuse patterns of SSL/TLS [S25] and Spring [S38] APIs. The research in [S38] implemented FSMs to monitor the program's authorization state for each type of misuse. Transitions between states occur when method calls are made to authorize the user or gain access to a critical resource.

\noindent\textbf{\textit{(iv) Code-based templates:}} Crafting language-, graph-, and FSM-based templates requires substantial domain knowledge and manual effort to specify the critical elements for API usage, their correlation, and modeling them as templates. An alternative solution is to automatically infer templates from sample security API use cases within the source code. In our review, one study [S2] used (insecure, secure) code pairs from prior research and compared their ASTs to identify edit operations. Vulnerability and repair templates were then generated based on a data-dependency analysis of ASTs and variable abstraction. The vulnerability template was used to detect misuse through pattern matching, while the repair template was used to generate customized fixes. Another study [S46] analyzed commits from 48 GitHub applications to manually identify misuses. ASTs were then used to identify required edit operations for fixes, which were clustered later based on their similarities. Finally, each cluster was generalized to vulnerability and repair templates to be used for detecting and fixing misuses. While code templates enable automatic pattern generation, they are inherently limited to known misuses reflected in existing code.

Table~\ref{tab:patt} summarizes descriptions, strengths, and weaknesses of pattern types and representation models, and Figure~\ref{fig:dist-heuristic}.a shows their distribution in primary studies.
\begin{table}[]
\centering
\caption{Pattern Inference categorizations with their descriptions, strengths, and weaknesses}
\label{tab:patt}

\footnotesize
\renewcommand{\arraystretch}{1}
\begin{tabularx}{\textwidth}{m{0.085\textwidth} m{0.27\textwidth} m{0.275\textwidth} m{0.27\textwidth}}

\hline
\textbf{Type} & \textbf{Description} & \textbf{Strengths} & \textbf{Weaknesses}\\
\hline

\multicolumn{4}{c}{\textbf{{Pattern Type}}} \\\hline
\textbf{Normal}& Patterns are inferred from correct uses of APIs, and any violation of these patterns is considered as misuse. &
$\bullet$ Limited number of patterns\newline
$\bullet$ Incomplete specification does not result in missed vulnerabilities&
$\bullet$ Susceptible to high false alarm rates
\\\hline
\textbf{Misuse}& Patterns are inferred from incorrect uses of APIs, and any matches with these patterns are considered as misuses.&
$\bullet$ Incomplete specification does not result in false alarms&
$\bullet$ Difficult to capture all possible patterns\newline
$\bullet$ Incomplete specification results in misuses being missed\\\hline

\multicolumn{4}{c}{\textbf{{Pattern Representation Model}}} \\\hline
\textbf{Hard-coded rules}&Patterns are defined as a set of rules.&
$\bullet$ Simple to design&
$\bullet$ Dependent on domain knowledge\newline
$\bullet$ Hard to extend to new misuses\\\hline
\textbf{Template}&Patterns are abstracted via a higher-level template.&
$\bullet$ Easier to extend to new misuses&
$\bullet$ Difficult to design a template from instances\\

\hline
\end{tabularx}

\end{table}

\vspace{5pt}
\noindent\textbf{C. Program Analysis Techniques:}
Our review identified three categories for dividing program analysis techniques based on their reliance on code execution: \textbf{\textit{(i) static analysis}}, \textbf{\textit{(ii) dynamic analysis}}, and \textbf{\textit{(iii) hybrid analysis}}. In the following, we investigate these methods and their adoption in the existing literature. 


\noindent\textbf{\textit{(i) Static Analysis:}} Static analysis involves examining (recovered) source code, binary code, or an intermediate representation of binary code without executing the application. This approach is also known as \textit{white-box} testing as it requires application code or its implementation details to identify misuses. Static analysis offers significant advantages: it is resource- and time-efficient, and it can achieve high code coverage. Additionally, developers can integrate static analysis tools into their daily workflow to identify misuse early in the development process.
However, static analysis also has limitations, such as high false alarm rates caused by infeasible misuses (that never occur at runtime) and failure to capture runtime misuses.

The main objective of static analysis is to determine the possible values of the parameter objects in a relevant API call and examine them against normal or misuse patterns to detect misuse. This is achieved through \textit{\textbf{data flow analysis}}, which typically uses \textit{program dependency graphs} or \textit{abstract syntax trees} to understand how data is used and manipulated within a program. 
There are two types of data flow analysis: \textit{\textbf{intra-procedural}} and \textit{\textbf{inter-procedural}}, depending on whether the interactions between different procedures or functions are considered or not. Most static approaches in our review rely on inter-procedural analysis, which enables the capture of more complex misuses. OAuthLint [S30], for instance, performed inter-procedural data flow analysis using Flowdroid~\cite{arzt2014flowdroid} to identify key elements for misuse patterns. Another tool based on inter-procedural analysis is CogniCrypt\textsubscript{SAST}~[S11], which was designed as a compiler for CrySL (a language-based template for normal use of crypto APIs; detailed in pattern representation models) to check Java source code for compliance with CrySL and generate code for common crypto tasks. Several studies [e.g., S36, S40, S58, S61] used CogniCrypt\textsubscript{SAST} to detect misuse of crypto APIs. There are also several studies [e.g., S28, S38, S59] based on intra-procedural analysis. For example, one study [S28] used data flow within cryptographic functions to identify paths taken by a parameter from its initial origin to its ultimate use within a function. 

To achieve more efficient misuse detection, several studies applied \textit{\textbf{program slicing}}, which simplifies the complexity of a program by removing parts of the code that are irrelevant to a specific analysis or task~\cite{xu2005brief}. This is accomplished by computing a set of program statements that affect (backward slicing) or are affected by (forward slicing) a given slicing criterion, which is typically an API parameter, based on data flow~[S1]. For instance, CryptoTutor~[S17] applied inter-procedural data flow analysis and program slicing to detect crypto misuses in Java code.
Similarly, CryptoLint~[S4] used inter-procedural backward slicing to track flows between crypto parameters and operations. Later, BinSight [S12] and CDRep~[S16] leveraged CryptoLint to examine the current state of crypto API usage in Android applications, with additional efforts towards source attribution~[S12] and repair~[S16]. Amandroid~[S44] also applied inter-procedural data flow analysis to assess the security state of Android apps in terms of data leaks, data injection, and improper use of crypto and SSL/TLS APIs. Program slicing was also used in the context of the Fingerprint API to classify applications into different security levels based on the state of their API use. This involved performing inter-procedural backward slicing to extract API parameters as features in a rule-based classification system [S37]. 

Although program slicing improves the efficiency of static analysis, it may lead to large memory and runtime overhead on massive-sized projects.  
To address this challenge, CryptoGuard [S1] proposed a trade-off between accuracy and scalability by performing on-demand slicing. This approach limits the analysis to methods that have the potential for security impact, effectively reducing the size of the code that needs to be analyzed.
Additionally, it utilized refinement algorithms to remove irrelevant language-specific elements and mitigate the high rate of false alarms in static analysis. Later, another study [S39] used CryptoGuard to detect misuse of crypto and SSL/TLS APIs in Android applications.

Another technique to minimize false alarms in static analysis is \textbf{\textit{symbolic execution}}~\cite {king1976symbolic} that executes a program by using symbolic values as inputs, rather than concrete values, and expressing the values of program variables as symbolic expressions of these inputs. Several studies, such as SSLDoc [S21] and TaintCrypt [S25], leveraged symbolic execution to statically detect security API misuse by creating program path traces that capture semantic information for each targeted API. Another study [S18] performed a simple variant of symbolic execution to extract crypto API sequences from Android applications, which were then used to learn probabilistic models to predict misuses. Additionally, some studies performed manual code analysis to detect crypto misuses [S35, S60, S61, S64, S66] and OAuth misuses [S48].

\vspace{3pt}
\noindent\textbf{\textit{(ii) Dynamic Analysis: }}
Dynamic analysis involves executing the code of an application and monitoring its behavior during runtime. It is also known as \textit{black-box} testing as it treats applications as black boxes and only considers the external behavior of an application at runtime. Dynamic analysis approaches do not usually produce false positives and can capture misuses occurring during runtime. However, dynamic analysis is resource-intensive and time-consuming, involving tasks such as installing, configuring, and testing, some of which may require human intervention [S23]. It also has limitations in terms of code coverage. There are two types of dynamic analysis for discovering software vulnerabilities, including API misuses: \textbf{\textit{active}} and \textbf{\textit{passive}}. \textit{Active dynamic analysis} involves intentionally attempting to exploit vulnerabilities or cause disruptions in a system. In contrast, \textit{passive dynamic analysis} focuses on collecting data and observing behavior without trying to cause harm. 

The passive dynamic analysis examines execution logs, the memory state of a program, or network traces to gain insight into its behavior. Considering the observed data, we have categorized passive dynamic approaches into \textbf{\textit{log}}, \textbf{\textit{memory}}, and \textbf{\textit{network analysis}}. \textit{Log analysis} involves collecting runtime information and execution traces and performing offline analysis after the execution is completed. While the offline analysis does not affect the application's performance [S10], it can generate large log files, creating an I/O bottleneck slowdown~[S13]. In our review, a few studies performed log analysis to detect misuse of security APIs. For example, one study [S10] examined logs that record parameters relevant to crypto API calls to find matches with some misuse patterns.  \textit{Memory analysis} was also utilized by some studies for misuse detection. For example, K-Hunt [S13] tracks memory buffers that store encryption keys to verify whether keys are generated and transmitted securely. It started with a lightweight dynamic analysis to gather runtime information required to locate memory buffers where crypto keys were stored. Meanwhile, several studies adopted \textit{network analysis} to detect misuses of APIs such as SSL/TLS and OAuth. One study [S52] evaluated the implementation of CSRF protection in OAuth transactions by checking the presence or absence of a \textit{state} variable in URLs.

In addition, some studies performed active dynamic analysis to verify the results obtained from network analysis. Active dynamic analysis can include a range of techniques, such as \textbf{\textit{penetration testing}}, which simulates a real-world attack on a running application to identify any misuses that could be exploited. It is considered the most effective approach to uncover exploitable misuses and avoid false alarms. One study [S50] manually analyzed the HTTP messages to capture the information flow of SSO credentials and detect potential misuse of OAuth. It further designed exploits to prevent manual inspection errors. Similarly, another study [S49] performed network analysis, followed by examining the feasibility of a CSRF attack to uncover exploitable misuses of OAuth.

\begin{table}[t]

\centering
\caption{Program analysis techniques with their descriptions, strengths, and weaknesses}
\label{tab:techs}

\footnotesize
\renewcommand{\arraystretch}{1}
\begin{tabularx}{\textwidth}{m{0.08\textwidth} m{0.27\textwidth} m{0.28\textwidth} m{0.275\textwidth}}

\hline
\textbf{Type} & \textbf{Description} & \textbf{Strengths} & \textbf{Weaknesses}\\
\hline

\textbf{Static}& Static analysis examines the application’s code against API usage constraints. &
$\bullet$ Doesn’t require program execution and is scalable to a large number of applications. &
$\bullet$ Applicable only to open-source projects\newline
$\bullet$ Suffers from high false alarm rate\\\hline
\textbf{Dynamic}& Dynamic testing executes the software and validates output or runtime information against API usage constraints. &
$\bullet$ Able to capture misuses occurring during runtime &
$\bullet$ Requires code execution \newline  
$\bullet$ Costly and not scalable to a large number of projects \\

\hline
\end{tabularx}

\end{table}

\vspace{3pt}
\noindent\textbf{\textit{(iii) Hybrid Analysis: }}
Attempts have been made to combine static and dynamic analyses in a \textbf{\textit{hybrid}} approach to leverage the strengths of both techniques and overcome their weaknesses. Table~\ref{tab:techs} provides a concise summary of the strengths and weaknesses associated with static and dynamic analysis techniques.
To mitigate the risk of false positives in the static analysis, some researchers proposed a hybrid approach that typically applies static analysis to identify potential misuses, followed by dynamic analysis to validate the results. Several studies [S8, S14, S19, S24, S57, S67] evaluated Android apps against a MitM attack to verify misuses reported by static analysis. 
Another study [S23] applied manual static analysis to find potential crypto misuses in Android apps and then performed dynamic memory analysis to examine the crypto libraries invoked during execution. This approach enables the detection of misuses that are feasible at runtime.
Another study [S69] used a combination of static and dynamic analyses to find Android applications that call the SafetyNet Attestation API during their execution. Next, it did a manual static analysis to find vulnerable applications with potential misuses, followed by bypassing the SafetyNet Attestation checks to confirm the misuses. 

Static analysis can also serve as a guide for dynamic analysis, reducing the time and memory consumption of dynamic analysis by pruning its exploration space.
Some studies [e.g., S19, S24, S67] employed a preliminary static analysis to detect misuses. They further used static analysis for method call graphs to identify the entry points that trigger the execution of vulnerable methods. These entry points were then used to generate inputs for running applications during dynamic analysis, resulting in a more efficient analysis with a reduced input space. 
Another study [S6] combined static and dynamic analysis techniques to detect crypto misuses in iOS apps. It first used static analysis to find the locations of crypto APIs and then monitored those API calls at runtime using \textit{API hooking} techniques. Misuses were detected by analyzing the execution logs, which record parameter values and other relevant information.
AuthDroid [S47] also adopted a hybrid approach to detect OAuth misuse in Android apps. It uses static analysis to extract the basic elements of OAuth (e.g., user-agent, the identity of SP) from the app, then uses a MitM proxy in dynamic analysis to find API misuses in the authentication process.
While the mentioned studies followed a static-dynamic approach in detecting misuses, one study [S26] has taken a different hybrid approach by first simulating a MitM attack to find vulnerable apps, and then manually performing static code analysis to identify the root causes of misuses.





\subsubsection{ML-based}
Our review found only 3 studies using ML-based algorithms to detect security API misuse. The main idea is to classify API usage instances within a given application as correct or incorrect using features reflecting the application's behavior. Below, we examine the feature engineering and classification components of these approaches.

\textbf{A. Feature Engineering:}
Three types of features were identified in the existing literature for building security API misuse detection models, which are \textit{\textbf{sequential, word-}}, and \textit{\textbf{graph-based}} features.

\noindent\textbf{\textit{(i) Sequential features:}} One study [S18] used API sequences representing both API orders and API arguments to learn probabilistic models for misuse detection. To this end, they used static analysis to extract possible traces for each reachable method from application binary files. Furthermore, they performed a simple variant of symbolic execution on each trace and then filtered traces of irrelevant APIs.

\noindent\textbf{\textit{(ii) Word-based features:}} Term frequency-inverse document frequency (tf-idf) is a common technique used in Natural Language Processing (NLP) to evaluate the importance of a term in a document or corpus. Recent advances in NLP have inspired many researchers to apply it to analyzing source code by considering it as natural-language text. In our review, one study [S45] extracted tf-idf from source codes to train a misuse detection model.

\noindent\textbf{\textit{(iii) Graph-based features:}} One study [S65] utilized graph-based features to analyze the usage of security APIs in source code. First, the source code was parsed to an AST and then modeled through graph embedding techniques, Bag of Graphs (BoG), and node2vec. These techniques are similar to word embedding techniques in NLP, where words are mapped to vectors based on how often they appear together in text. Similarly, BoG generates a collection of graph bag items representing elements or sub-graphs within a graph. These items are then used to construct a vector representation that captures the local attributes and relationships of the original graph. Node2vec is another graph embedding technique that extracts features from graphs, utilizing a flexible neighborhood sampling strategy.

\textbf{B. Classification:} In one study [S18], two probabilistic models, Hidden Markov Model (HMM) and n-gram, were trained using both secure and insecure API sequences. Notably, the labels for these sequences were generated using an existing tool, CogniCrypt\textsubscript{SAST} [S11], designed for detecting misuses of cryptography APIs. The trained models were employed to predict the probability of a given API sequence being secure. An API sequence was considered insecure if its probability fell below a pre-defined threshold. The study also addressed the problem of identifying misuse locations within an insecure sequence by using a distance measure based on the probability of an API misuse at possible locations. 
In another study [S45], code snippets with the usage of security APIs were mined from SO and then classified using a Support Vector Machine (SVM) model. A small set of extracted code snippets was manually labeled to build the training dataset. Similarly, the approach proposed in [S65] used SVM, but trained a separate classifier for each misuse type, enabling the model to identify both the presence and type of API misuse. The classifiers were trained using labeled data containing correct and incorrect API use instances from two existing benchmarks for evaluating crypto misuse detection techniques, namely datasets by Braga et al. ~\cite{braga2017practical} and Fischer et al.~\cite{fischer2019stack}.

\section{RQ4: Evaluation methods}\label{sec:eval} 
Our review identified 25 studies that performed experiments to assess the performance of misuse detection techniques. Out of these, 5 studies [S31, S32, S34, S41, S42] evaluated the performance of various static analysis tools in detecting crypto and SSL/TLS misuses. Following, we discuss various metrics, benchmarks, and strategies adopted for evaluation.

\subsection{Evaluation Strategies}\label{sec:evaluation}

In our review, various techniques were used to evaluate the performance of misuse detection models. We noticed only seven studies designed experiments using \textbf{\textit{public benchmarks}} (\S~\ref{sec:benchs}). Existing benchmarks are typically limited in terms of scale and diversity of test cases. To address this issue, one study [S34] explored the \textbf{\textit{automatic generation of test cases}} using mutation operations. It generated over 20,000 test cases, which were used to evaluate several crypto misuse detection tools and identify their flaws, such as failure to detect insecure algorithms provided in lowercase.
In addition, 19 primary studies conducted \textit{\textbf{case studies}} or \textbf{\textit{manually analyzed}} a subset of their dataset or a subset of reported misuses to verify their results. For instance, one study [S10] randomly selected 150 Android apps out of 1,780 analyzed apps to validate their findings, and another study [S58] randomly sampled 157 misuses and manually verified them to gain a deeper understanding of common false positives. 

The study [S45] created a dataset for 5-fold cross-validation by manually labeling a collection of security-related code snippets from SO as either secure or insecure that were used for evaluating its proposed ML-based algorithms. Unlike the study [S45], the study [S18] relied on an \textbf{\textit{existing tool}}, CogniCrypt\textsubscript{SAST} [S11], to label crypto API use cases in Android applications and provide a labeled dataset for training, validating, and testing purposes of its ML algorithms, which makes the results dependent on the performance of the employed tool.	
 
Another evaluation technique, adopted by 3 studies, involves \textbf{\textit{executing attacks}} to validate the results and identify exploitable misuses. For instance, one study [S6] executed two ethical attacks on two applications and successfully retrieved personal information encrypted and transmitted over the network. Meanwhile, 16 studies disclosed misuses identified in real-world projects, some of which analyzed the feedback they received from developers. This analysis provided valuable insights into developers’ requirements for misuse detection tools, disregarded in existing approaches. 

Lastly, we found 3 studies that conducted \textbf{\textit{user studies}} to evaluate the usability of tools with warning messages and suggestions for fixes. These studies involved 39 developers [S5], 8 developers [S42], and 53 developers [S59] and showed that security advice could improve the usage of crypto APIs in users’ codes. More importantly, they highlighted the need for detailed and specific solutions that are comprehensible and feasible for developers.

\subsection{Evaluation Benchmarks} \label{sec:benchs}

\begin{figure}

\begin{minipage}[c]{0.8\textwidth}
\centering
\captionof{table}{Public benchmarks for evaluating crypto misuse detection techniques}
\label{tab:benchs}
\footnotesize
\setlength{\tabcolsep}{4pt}

\begin{tabular}{p{0.2\textwidth} C{0.1\textwidth} C{0.02\textwidth} p{0.5\textwidth} p{0.06\textwidth}}

\hline
\textbf{Benchmark} & \textbf{Size} & \textbf{R/S\textsuperscript{1}} & \textbf{Description} & \textbf{Ref}\\
\hline
\textbf{CryptoAPI-Bench} (2019) [S1]\cite{cryptoapi-bench}  & 181\newline test cases & S & Benchmark for evaluating crypto misuse detectors containing 45 basic and 136 complex test cases with crypto API misuses, false positive tests, and correct API uses & S1, S2, S31, S32
\\
\hline
\textbf{CryptoAPI-Bench*} (2021) [S10]  & 198\newline test cases & S & CryptoAPI-Bench with further cases suitable for assessing dynamic approaches, totally consisting of 157 crypto misuse cases, and 41 normal test cases & S10
\\
\hline
\textbf{ApacheCryptoAPI-Bench} (2020) [S32]\cite{Apachecryptoapi-bench}&  120\newline test cases & R & Ten real-world Apache projects including 79 basic test cases and 42 advanced test cases, suitable for assessing the scalability of misuse detection approaches
 & S32
\\
\hline 
\textbf{Braga et al.'s dataset} (2017)~\cite{braga2017practical}  & 384\newline test cases & S & Contains 202 misuses (positive cases) and 182 normal uses (negative cases) for Java Cryptography Architecture & S41, S65
\\
\hline
\textbf{Fischer et al.'s dataset} (2019)~\cite{fischer2019stack}&  16,346\newline test cases & R & 6,246 secure cases and 10,100 insecure cases for the use of crypto API adopted from code snippets available at SO posts
 & S65
\\
\hline
\textbf{MUBench} \newline(2016)~\cite{amann2016mubench}\cite{MUBench} & 21\newline apps& R & Benchmark for evaluating API-misuse detectors containing instances of crypto API misuses from 62 Java programs & S31
\\
\hline
\textbf{OWASP} \newline(2021)~\cite{owasp} & 975 \newline programs & R & Java test suite designed for evaluating vulnerability detectors, containing 477 programs with labeled misuses of security APIs and 498 programs with correct uses & S31
\\
\hline
\textbf{DroidBench} \newline(2015)~\cite{DroidBench}  & 21\newline apps& R &  Benchmark apps for evaluating the performance of static information-flow analysis of Android apps, including crypto misuse test cases & S44
\\
\hline
\textbf{ICC-Bench} \newline(2017)~\cite{ICC-Bench}& 24\newline apps& R & Benchmark apps for evaluating the performance of static analysis to detect inter-component data leakage problem of Android apps, including crypto misuse test cases & S44
\\
\hline
\end{tabular}

        \begin{flushleft}
        \footnotesize
        1. Real or Synthetic 
        \end{flushleft}            
  \end{minipage}
\hfill
\begin{minipage}[c]{0.19\textwidth}
\includegraphics[width=\linewidth]{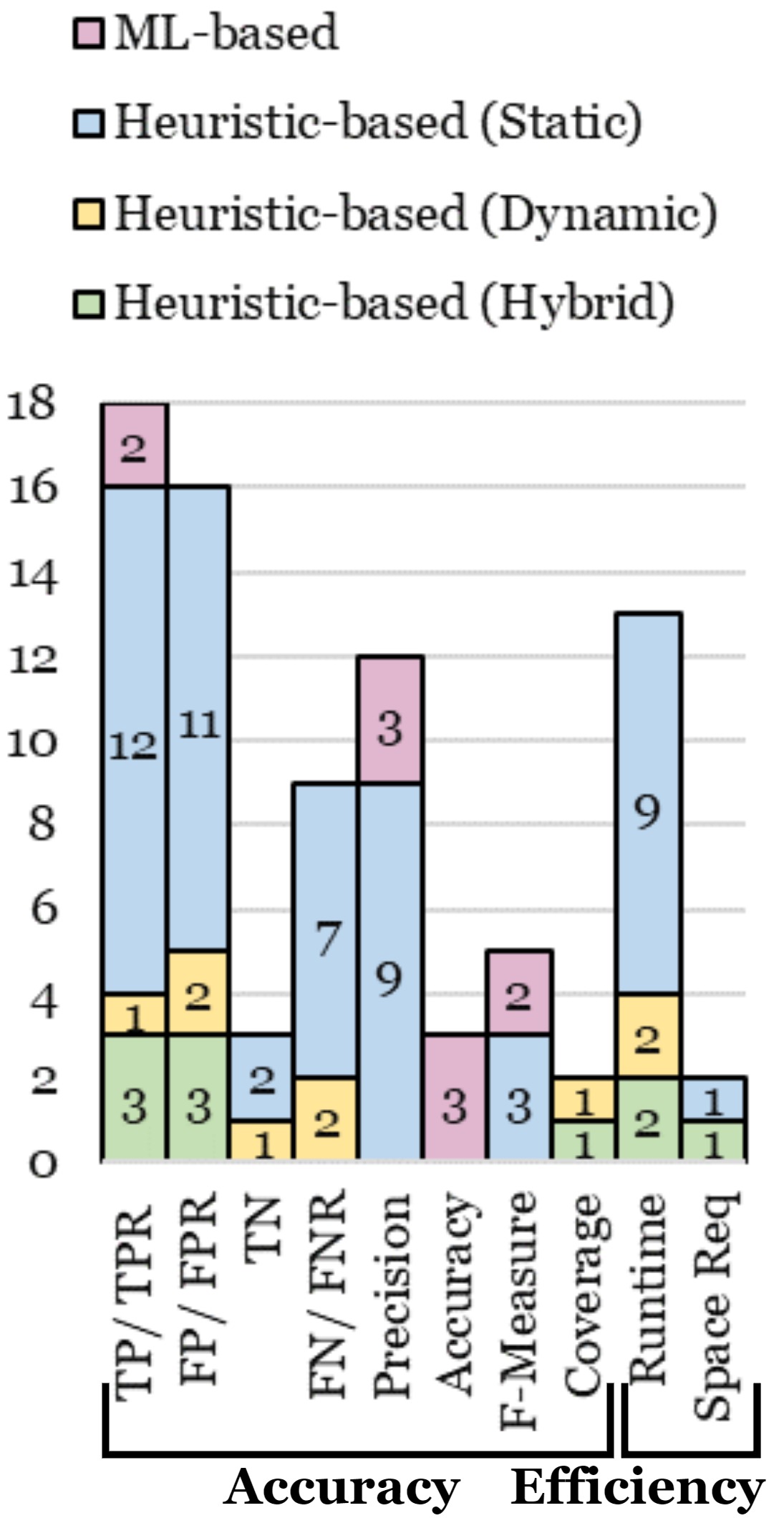}
\vspace{3pt}
\caption{Evaluation metrics Distribution categorized by\\ detection techniques}
\label{fig:dist_metrics}
\end{minipage}

\end{figure}

Benchmarks are critical for evaluating detection techniques and identifying their strengths and weaknesses. Unfortunately, there is an inadequate number of publicly available benchmarks, and all of them are limited to test cases for some misuses of Java crypto and SSL/TLS APIs. Table~\ref{tab:benchs} lists 9 public benchmarks commonly used by researchers. Among them, the first five benchmarks were specifically created to evaluate and compare the performance of misuse detection approaches for security APIs. 

\textit{\textbf{CryptoAPI-Bench}} includes synthetic source code examples with crypto API misuses, false positive tests, and correct API uses. It offers both basic test cases and advanced test cases that involve more complex scenarios. CryptoAPI-Bench was designed to assess static tools. \textbf{\textit{CryptoAPI-Bench*}}~[S10] is an extension of CryptoAPI-Bench with additional cases suitable for evaluating dynamic approaches. CryptoAPI-Bench is not suitable for evaluating the scalability of a tool, as all test cases are lightweight by design. To address this limitation, another study [S32] created \textit{\textbf{ApacheCryptoAPI-Bench}} using 10 real-world Apache projects that are complex programs with numerous and lengthy code files. This benchmark is therefore appropriate for assessing the scalability and applicability of existing approaches to real-world applications. 

Two additional benchmarks for evaluating crypto misuse detection techniques are \textbf{\textit{Braga et al.'s}}~\cite{braga2017practical} and \textbf{\textit{Fischer et al.'s}}~\cite{fischer2019stack} datasets that contain labeled instances of secure and insecure use of the Java Cryptography Architecture API. Braga et al.'s dataset consists of synthetic Java source codes, while the latter consists of real-world code snippets collected from SO. Both datasets were used by the study [S65] to train and test an ML-based detection technique. The last four benchmarks have been designed to evaluate API misuse or vulnerability detectors and include some test cases for evaluating crypto misuse detection techniques as well.

\subsection{Evaluation Metrics}

We have identified 10 evaluation metrics that are commonly used to measure the performance of security API misuse detection techniques. These metrics are grouped into two categories: \textit{\textbf{detection effectiveness}} and \textbf{\textit{computation efficiency}}. 
Metrics for detection effectiveness are typically calculated using \textit{True Positive (TP), False Positive (FP), True Negative (TN)}, or \textit{False Negative (FN)} values. While detecting all misuses is crucial, having a high number of false alarms can be highly time-consuming and burdensome for developers. Hence, the primary objective of misuse detection is to maximize the \textbf{\textit{True Positive Rate (TPR)}} or \textbf{\textit{Recall (R)}}, while minimizing the \textit{\textbf{False Positive Rate  (FPR)}}. These two metrics are the most commonly used. Other metrics, such as \textbf{\textit{True Negative Rate (TNR)}}, \textbf{\textit{False Negative Rate (FNR)}}, \textbf{\textit{Precision (P)}}, \textbf{\textit{Classification Accuracy}}, \textbf{\textit{F-Measure}}, and \textbf{\textit{Coverage}} have also been used by researchers to measure the effectiveness of misuse detection.

Several primary studies also considered the computation efficiency of detection techniques, which was measured using the \textbf{\textit{runtime}} and \textbf{\textit{space}} complexity required for misuse detection. Evaluating computation efficiency is crucial in demonstrating the suitability of these techniques for real-world applications.
Classification accuracy was used only by ML-based detection techniques, and coverage measurement is exclusive to dynamic analysis, as it measures the proportion of the program code that has been executed during testing. Figure~\ref{fig:dist_metrics} illustrates the distribution of the identified evaluation metrics that are categorized by detection technique.

\section{Open Research Challenges and Future Directions}\label{subsec:open_issues}

\begin{figure}[t]
    \centering\includegraphics[width=\textwidth]{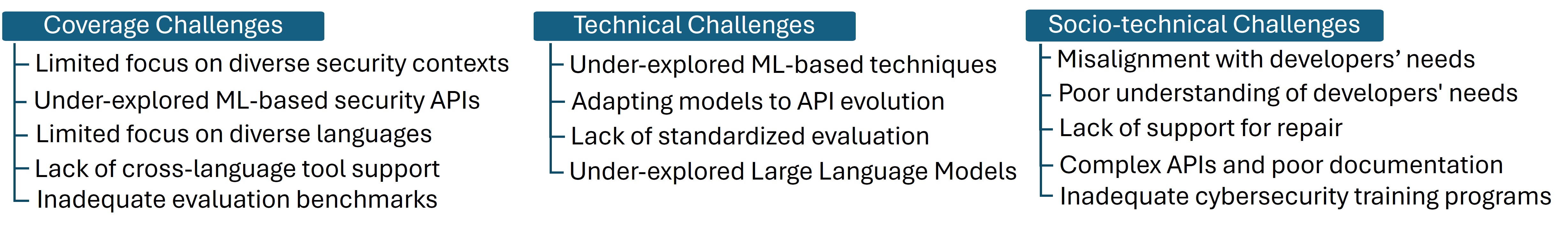}
    \caption{Open issues in enhancing and evaluating security API use}
    \label{fig:challenges}
\end{figure}

Our review identified 14 critical open issues in enhancing and evaluating security API use, categorized as coverage, technical, and socio-technical issues (detailed in Figure~\ref{fig:challenges}). In the following, we discuss these challenges and propose recommendations for future research to address them.

\subsection{Coverage Challenges} This category includes challenges arising from the limited scope of existing research as discussed below.

\subsubsection{Limited focus on diverse security contexts}

Although the focus of primary studies on crypto and SSL APIs (83\% of reviewed studies) highlights their crucial role in secure software development, it raises a warning alarm about potential vulnerabilities caused by misusing security APIs that receive inadequate or even no attention from the research community. Meng et al.'s analysis of SO posts~\cite{meng2018secure} revealed that Spring Security is the most popular option among Java developers in secure coding practices. However, our review found only one study dedicated to this overly complicated and poorly documented API. However, research suggests that misusing other security APIs, including those designed to store and access sensitive information, can have significant security implications~\cite{bosu2017collusive, nan2018finding, zuo2019does}. 
As a result, developing effective mechanisms to detect and prevent API misuse in these contexts is critical.

\subsubsection{Under-explored ML-based security APIs}
Novel security APIs have been developed to facilitate the integration of applications with cutting-edge security technologies, including various ML-based authentication schemes such as facial recognition, vocal recognition, and iris-based authentication~\cite{kunda2021survey}. While their adoption is increasing in various applications, research on their usability and usage patterns remains limited. 
The reliance of these APIs on probabilistic ML models introduces unique challenges in detecting their misuse~\cite{wan2021machine,wan2022automated, wei2024demystifying}. Unlike traditional APIs, where misuse might lead to explicit errors, misuse of ML-based APIs can result in suboptimal or incorrect predictions that deviate from human judgment~\cite{wan2022automated}. 
This makes the analysis and identification of misuse significantly more complex.
Therefore, further research is imperative to devise and develop methodologies to identify and mitigate the misuse of ML-based security APIs.

\subsubsection{Limited focus on diverse programming languages}

Over 70\% of the studies reviewed focused on Java, reflecting its extensive adoption and complex API design. Nonetheless, it is essential to recognize the significance of exploring and addressing existing challenges for using APIs in other widely used programming languages such as Python and C. Several primary studies on non-Java languages indicate that developers in other languages are also likely to make mistakes when using security APIs [e.g., S6, S9, S13]. This underscores the critical need for developing tools across diverse programming languages, with essential support from the research community.
It is highly recommended that researchers broaden their research scope to include non-Java languages in their analysis of security API misuse.

\subsubsection{Lack of cross-language tool support}
Along with the need to develop tools for various languages, there is a need to develop cross-language tools. A study [S20] analyzing MicroPython projects suggests that developers working in the embedded domain often use C~language to implement cryptography operations. This indicates the need for tools with cross-language analysis capability to track program information across multiple programming languages. Furthermore, Meng et al.'s analysis of SO posts~\cite{meng2018secure} has shed light on the challenge of cross-language data handling of cryptography APIs, where developers struggle to encrypt data in one language (e.g., PHP or Python) and decrypt data in another language (e.g., Java). To address these issues, future research should address the development of cross-language approaches to ensure secure software development across different programming languages.

\subsubsection{Limited availability and scope of evaluation benchmarks}

Benchmarks are crucial to compare and evaluate the performance of proposed approaches and tools for security API misuse detection and to identify the areas for further improvement.  However, our review identified only 9 publicly available benchmarks, out of which only 5 were specially designed for security API misuses. All these benchmarks are limited to test cases for some misuses of Java crypto or SSL/TLS APIs.  Notably, three out of these five benchmarks are synthetic, rendering them ineffective for evaluation in real-world settings. Thus, there is a desperate need to expand the scope of benchmarks to cover a broader range of security APIs, real-world misuses, programming languages, and software platforms suitable for evaluating tools developed for diverse security APIs in real-world scenarios. On the other hand, existing benchmarks are mostly limited to test cases for evaluating static tools, which renders them unsuitable for assessing the performance of dynamic analysis tools [S10]. Therefore, benchmarks need to come with test cases for evaluating dynamic analysis tools. It is also crucial to continuously update existing benchmarks to incorporate new misuses and the constant evolution of APIs. Furthermore, we recommend versioning benchmarks to address issues such as concept and temporal drift.

\subsection{Technical Challenges} Our review identified five technical-related open issues, detailed below.

\subsubsection{Limited research on the application of ML-based techniques}

Our SLR identified a critical gap: \textit{a scarcity of state-of-the-art ML-based models} for identifying security API misuse.  Over 95\% of the reviewed studies rely on traditional heuristic-based approaches that require significant domain knowledge and result in labor-intensive, time-consuming, and error-prone processes. Manual analysis of identified misuses in one study [S58], using heuristic-based CrySL [S11], also revealed several false positives due to incorrect specifications defined in CrySL.
Additionally, these methods often rely on hard-coded rules, making them difficult to adapt to emerging misuses to keep pace with the rapidly evolving security threat landscape. 
The main obstacle to adopting ML models in this domain is the scarcity of large-scale, labeled datasets. While some studies have manually analyzed and verified identified misuses, none have publicly shared the results of their analysis. We strongly encourage researchers to openly share their findings on security API usage analysis. This would not only support validation and future research but would also significantly benefit the development of large-scale labeled datasets, crucial for both training ML models and establishing evaluation benchmarks.





\subsubsection{Under-explored application of Large Language Models} In recent years, the rise of Large Language Models~(LLMs) has significantly influenced the field of software engineering~\cite{hou2023large}. Their capability to comprehend and generate code has demonstrated remarkable performance across various tasks like code completion~\cite{ross2023programmer}, explanation~\cite{nam2024using}, and repair~\cite{xia2023automated}. This suggests LLMs as potential tools for addressing security API misuse, a critical yet under-explored area. While a study by Wei et al.~\cite{wei2024demystifying} using ChatGPT demonstrated promising results in detecting and fixing API misuse, its scope is limited to a specific set of ML APIs. The research by Mousavi et al. ~\cite{mousavi2024investigation} revealed limitations in LLMs' ability to generate secure code for security APIs. Further research is crucial to gain an in-depth understanding of LLMs' strengths and limitations in addressing security API misuse. Additionally, future investigations should explore avenues to enhance security awareness in LLMs, including strategies such as security-enhanced training, integrating security considerations during fine-tuning or prompt tuning, and adapting LLM generation to reinforce adherence to established security best practices.

\subsubsection{Challenges in adapting models to API evolution}

The ever-changing nature of APIs presents a significant challenge for misuse detection models. As APIs adapt to new requirements and security concerns, detection models must constantly keep pace to identify new misuse patterns arising from these changes. However, the rapid evolution often outpaces existing models' ability to detect new forms of misuse. In our analysis, we noted the deprecation of two APIs, Fingerprint and SafetyNet Attestation, yet no solutions have addressed adapting misuse detection models to their alternatives, Biometrics \cite{biometrics} and Play Integrity \cite{play_integrity} APIs. Zhong and Meng’s recent work~\cite{zhong2024compiler} takes a step toward addressing this issue by introducing a compiler-directed approach that automatically updates callsites when compilation errors indicate that a previously valid API usage has become incompatible. This offers a promising direction for bridging the gap between evolving APIs and misuse detection models when changes result in compilation failures. However, many evolution scenarios—particularly in the security domain—do not manifest as compilation errors. Future research should therefore extend automated migration techniques to handle such cases and integrate these techniques with misuse detection models to ensure their continued effectiveness as APIs evolve. Another significant issue is the overreliance on hard-coded rules (over 95\% of studies), which limits the ability to detect emerging misuse patterns. We therefore recommend that future work investigate ML-based models and efficient incremental learning algorithms to address concept drift as APIs continue to evolve.

\subsubsection{Lack of a standardized evaluation framework}
Pendlebury et al. \cite{pendlebury2019tesseract} identified two critical sources of bias that inflate experimental results in ML-based malware classification: spatial bias and temporal bias. Spatial bias stems from unrealistic assumptions about the goodware-to-malware ratio within training and testing data. Temporal bias occurs due to unrealistic time splits within datasets, where future knowledge is included in the training data.
Their proposed evaluation framework addresses these issues by incorporating spatial and temporal constraints. This concern extends beyond malware classification to any machine learning model in the security domain that deals with constantly evolving threats and time-sensitive data, including ML-based security API misuse detection. Furthermore, heuristic-based approaches should also be tested under conditions that reflect actual usage scenarios and the evolution of APIs over time. However, no reviewed studies explicitly considered such biases. To ensure accurate assessment of misuse detection in realistic settings, the research community must prioritize integrating spatial and temporal constraints into experimental design. This specifically requires augmenting datasets with accurate timestamps and obtaining realistic ratios between normal and misuse cases.




\subsection{Socio-technical Challenges} Despite being the primary users of APIs and security tools, developers are often overlooked in both API and tool development. This section discusses five key socio-technical challenges in security API misuse detection.

\subsubsection{Inadequate models to meet developers' expectations}

Sixteen studies disclosed misuses identified in real-world projects, some of which [e.g., S1, S10, S31] reported the feedback received from developers. These studies revealed \textit{inadequacies of current tools in meeting the expectations and requirements of developers}, indicating the need for adapting these tools based on user feedback and preferences. In certain cases, developers acknowledged the existence of misuses (71\% [S21], 52\% [S9], 20\% [S29] of reported misuses) and attempted to address them. However, in some cases, they disregarded the identified misuses, believing them to be irrelevant. These misuses were often found in non-sensitive contexts such as security-irrelevant pieces of code, test cases, or archived code [S10, S31]. Additionally, in some cases, developers rejected misuses without concrete exploit demonstrations [S1, S10, S31]. To better align with developers' requirements, practitioners should consider developing detection tools capable of differentiating between security-relevant and security-irrelevant contexts and demonstrating concrete exploit examples for identified misuses. By addressing the human-centric aspects of misuse detection models, tool developers can enhance the usability, effectiveness, and adoption of these tools among developers.



\subsubsection{Lack of understanding of developers' needs}
Current research on misuse detection models overlooks \textit{understanding developers' needs}. Our review identified a significant gap in this area, with only three studies [S5, S42, S59] evaluating the usability of their tools in assisting developers. To fill this research gap, future research needs to focus on understanding and identifying developers' requirements and expectations from misuse detection models. We recommend conducting user studies to gather insights directly from developers to achieve a more in-depth understanding of their needs. By understanding and analyzing these requirements, researchers can craft human-centric strategies for integrating user feedback into the design and development phases of misuse detection tools.

\subsubsection{Lack of support for repair}

Developers may face challenges while trying to fix identified misuses, potentially introducing new mistakes [S42]. 
An analysis of vulnerability disclosures, reported by the reviewed studies, reveals instances where developers acknowledged misuses but struggled to resolve them due to several limitations. Operational constraints, like maintaining backward compatibility, can restrict their ability to implement necessary fixes [S1]. Additionally, existing tools often provide inadequate repair guidance, lacking the detailed information required for repair [S31]. Further, the inherent complexity of implementing secure solutions poses a significant challenge for developers [S31]. To address these challenges, misuse detection tools should be complemented with comprehensible, detailed, and actionable fixing suggestions.
While some studies offer general repair guidance for specific misuse types [S5, S59, S63], they lack customized fixes for a given vulnerable program. There are only four studies [S2, S16, S17, S46] for the automated generation of customized fixes for crypto misuses. Existing tools, however, are still inadequate in assisting developers with accurately correcting misuses [S31]. Future research should explore methods for providing detailed and customized suggestions and streamlining automated repair solutions.

\subsubsection{Inadequate API usability and poor documentation}

The intricate designs of security APIs create a significant barrier to understanding and proper implementation, often overwhelming developers with a complex set of programming options, numerous parameters, return values, and their security implications~\cite{green2016developers}. This complexity is further compounded by poorly configured APIs with insecure defaults, like JCA's default use of the \texttt{\small ECB} mode for \texttt{\small AES} encryption \cite {kruger2019crysl}. Another critical concern arises from inadequate or poor documentation that fails to provide explicit examples of proper usage or even includes API misuses, as seen within documentation by some OAuth API providers~\cite{shernan2015more}. This lack of clear guidance forces developers to rely on potentially unreliable sources like forum posts, which can lead to copying and pasting incorrect suggestions that contain misuse instances~\cite{fischer2017stack}. 
To address these challenges, there should be a focus on user-centric approaches to security API design. Security API development should prioritize clarity and ease of use while ensuring secure configurations by default and providing well-written documentation that streamlines the effective use of security APIs even for novice developers. To achieve this, we recommend that researchers investigate and identify usability issues in existing security APIs and devise effective mitigation strategies.


\subsubsection{Lack of effective cybersecurity training programs}

Inadequate cybersecurity training for developers is a major contributor to security API misuse~\cite{assal2018security}. Meanwhile, the threat landscape for security APIs is continuously evolving, with new attack techniques and vulnerabilities being discovered regularly. Hence, ongoing cybersecurity education and training for developers is crucial.  However, research to support effective training programs remains limited~\cite{espinha2020sifu, gasiba2020design, gasiba2021cybersecurity}. Therefore, future research should focus on identifying and overcoming barriers to the adoption of existing developer security training programs. The research also needs to explore innovative technologies, including virtual reality, gamification, and interactive online platforms for delivering engaging and effective security training~\cite{lampropoulos2024virtual}. Additionally, given the critical role of security APIs in software security, their secure use should be a top priority in training programs. Moreover, organizations must prioritize investments in developer training programs and ensure that their developers receive regular updates on cybersecurity best practices.

\color{black}
\section{Threats to Validity}\label{sec:threats}

We followed the guidelines outlined in study~\cite{slr_guidelines} to design and conduct our SLR. However, there might still be some biases due to the author's expertise and different perspectives. We took necessary steps to minimize the impact of potential biases, including those in the study selection and review process, which are elaborated upon below:

\noindent\textbf{\textit{Search Strategy}:} One of the common threats to the validity of an SLR is the possibility of missing relevant studies. To minimize this risk, we utilized Scopus, which is the most comprehensive search engine and largest indexing system~\cite{slr_triet_vul, roland}, and supplemented it with the two most frequently used digital libraries, IEEE Xplore and ACM Digital Library~\cite{digital_lib}. We also conducted a series of pilot searches to establish a search string that would retrieve relevant papers already known to us. In addition, we employed both forward and backward snowballing techniques to locate any other relevant papers that might have been missed by the search string.

\noindent\textbf{\textit{Selection process}:} The potential for subjective bias in the selection of studies cannot be ruled out, as it could be influenced by the author’s subjective judgment. To address this concern, we carried out a rigorous and well-defined multi-step process (detailed in \S~\ref {subsec:study_selection}) with clear inclusion and exclusion criteria. We also established specific guidelines to exclude low-quality papers. At every stage of the selection process, we carefully deliberated and addressed any uncertainties through discussions among all the authors to minimize the risk of selection bias. 
    
\noindent\textbf{\textit{Data Extraction and Synthesis}:} Human errors and author biases during data extraction, analysis, and interpretation can impact the accuracy of the results and findings. To mitigate this issue, a data extraction form was developed and refined to ensure the collection of adequate and consistent information for answering the research questions. We also conducted fortnightly meetings among all the authors to review and verify the synthesis and interpretation of our quantitative and qualitative analysis. Any disagreements were discussed and resolved collaboratively before finalizing our responses to the RQs.

\section{Conclusion} \label{sec: concl}

Security APIs play a crucial role in secure software development. Prior studies have shown developers often misuse security APIs, leading to costly software vulnerabilities. Thus, misuse detection for security APIs has gained significant attention from the research community for ensuring software security. However, the existing literature on the topic is dispersed, and a systematic review was necessary to identify the state-of-the-art approaches and highlight areas that require further exploration. This study presents our research effort aimed at systematically reviewing and rigorously analyzing the literature on misuse detection for security APIs. To the best of our knowledge, this SLR is the first attempt to systematically review the literature on this topic. We have provided an organized evidence-based body of knowledge to enrich this domain by identifying security APIs, their potential misuses, detection techniques, and evaluation methods. In conclusion, based on a comprehensive analysis of 69 primary studies, we identified the following key trends in security API misuse detection research:
\begin{enumerate}[label=\arabic*)]
    \item We identified 6 security APIs examined for misuse detection, namely cryptography primitives (crypto), SSL/TLS, OAuth, Fingerprint, Spring, and SafetyNet Attestation. Most studies focused on crypto and SSL/TLS, highlighting the need to explore this topic for other security APIs.
    \item We identified a total of 30 distinct types of security API misuses. The primary studies mainly focused on analyzing Android apps, and the most commonly reported misuses were using insecure crypto algorithms for crypto APIs, improper certificate validation for SSL APIs, lack or misuse of the \textit{state} parameter for OAuth APIs, and lack or misuse of cryptography for Fingerprint APIs. 
    \item We proposed a taxonomy consisting of heuristic-based and ML-based approaches for misuse detection techniques. Most studies relied on heuristic-based approaches, with 42 studies based on static analysis, 9 studies using dynamic analysis, and 13 studies using a hybrid approach. We found only 3 studies using ML to address misuse detection. Our findings suggest the need to explore the application of ML, DL, and NLP techniques in this area.
    \item We identified 11 metrics for evaluation, grouped into accuracy and efficiency categories. We found only five public benchmarks, particularly designed for security API misuse, which are limited to test cases for crypto and SSL/TLS misuses. These findings highlight the need for further research and development of more diverse benchmarks to facilitate the evaluation of misuse detection techniques for security APIs.
\end{enumerate}

Overall, our SLR offers valuable insights for both researchers and practitioners. Researchers can deepen their understanding of existing work and identify areas for their future studies. Practitioners benefit in two ways - firstly, by using the findings to develop more effective tools for detecting security API misuse and secondly, by gaining insights into potential misuses, enabling them to avoid them in their development processes. Ultimately, this review contributes to the broader goal of promoting secure software development.

\vspace{-5 pt}

\begin{acks}
The work has been supported by the Cyber Security Research Centre Limited, whose activities are partially funded by the Australian Government’s Cooperative Research Centres Programme.
\end{acks}

\vspace{-5 pt}

\bibliographystyle{unsrtnat}
\bibliography{sample-base}

\end{document}